\documentclass[english]{article}
\usepackage[T1]{fontenc}
\usepackage[latin9]{inputenc}
\usepackage{graphicx}
\usepackage{esint}
\usepackage{subscript}

\usepackage{hyperref}
\usepackage{cleveref}

\makeatletter

\newcommand{\lyxaddress}[1]{
	\par {\raggedright #1
	\vspace{1.4em}
	\noindent\par}
}

\usepackage{babel}

\makeatother

\usepackage{babel}
\begin{document}

\title{On the X-ray scattering pre-peak of linear mono-ols and the related
micro-structure from computer simulations}

\author{Martina Po\u{z}ar$^{1}$, Jennifer Bolle$^{2}$, \\
 Christian Sternemann$^{2}$\thanks{christian.sternemann@tu-dortmund.de}
and Aurélien Perera$^{3}$ \thanks{aup@lptmc.jussieu.fr}}
\maketitle

\lyxaddress{$^{1}$University of Split, Faculty of Sciences, Ru\d{j}eraBo\v{s}kovi\'{c}a
33, 21000, Split, Croatia}

\lyxaddress{$^{2}$Fakult\"at Physik/DELTA, Technische Universit\"at Dortmund, D-44221
Dortmund, Germany}

\lyxaddress{$^{3}$Sorbonne Universit\'e, Laboratoire de Physique Th\'e orique de
la Matière Condens\'e e (UMR CNRS 7600), 4 Place Jussieu, F75252, Paris
cedex 05, France.}
\begin{abstract}
The X-ray scattering intensities $I(k)$ of linear alkanols OH(CH$_{2}$)$_{n-1}$CH$_{3}$,
obtained from experiments (methanol to 1-undecanol) and computer simulations
(methanol to 1-nonanol) of different force field models, are comparatively
studied, particularly in order to explain the origin and the properties of the 
scattering pre-peak in the $k$-vector range $0.3\mathring{A}^{-1}-1\mathring{A}^{-1}$.
The experimental $I(k)$ show two apparent features: the pre-peak
position $k_{P}$ decreases with increasing $n$, and more intriguingly,
the amplitude $A_{P}$ goes through a maximum at 1-butanol ($n=4$).
The first feature is well reproduced by all force field models, while
the second shows a strong model dependence. The simulations reveal
various shapes of clusters of the hydroxyl head-group,
from $n>2$. $k_{P}$ is directly related to the size of the \emph{meta-objects}
corresponding to such clusters surrounded by their alkyl tails. The
explanation of the $A_{p}$ turnover at $n=4$ is more involved. The
analysis of the atom-atom structure factors indicates that the pre-peaks
arise from an incomplete cancellation between two types of contributions
from the various atom-atom structure factors: positive contributions
from atoms part of the head group (essentially hydroxyl group), and
a mostly negative contributions from the cross head/tail correlations.
The former dominate for small alkanols ($n<3$), while the latter
gain importance from $n=4$, which explains the pre-peak amplitude
turnover. The flexibility of the alkyl tails tend to reduce the cross
contributions, thus revealing the crucial importance of this parameter
in the models. Force fields with all-atom representation are less
successful in reproducing the pre-peak features for smaller alkanols
$n<6$, possibly because they blur the charge ordering process since
all atoms bear partial charges. The analysis clearly shows that it
is not possible to obtain a model free explanation of the features
of $I(k)$
\end{abstract}

\section{Introduction}

Radiation scattering is generally considered to provide the most direct
insight into the spatial microscopic structure of liquids, complementary
to thermodynamic or spectroscopy techniques. Linear mono-ols have
been investigated by X-ray and neutron scattering techniques since
the 1930s, and the most remarkable known feature is the presence of
a pre-peak of the scattering intensity $I(k)$, in the $k$-vector
region 0.3-1\,$\mathring{A}$ $^{-1}$ \cite{ExpScattOldAlc2,ExpScattOldAlc},
in addition to the usual main peak around $k\approx$1.4-2\,$\mathring{A}$ $^{-1}$.
The main peak can be interpreted through $\sigma=2\pi/k$ as corresponding
to the mean atomic size $\sigma\approx\,$3-4\,$\mathring{A}$, such
as in the X-ray or neutron diffusion in simple atomic liquids \cite{EXP_Kruh_diffraction}.
There has been several investigations of the origin of the pre-peak,
which is now generally interpreted as corresponding to the existence
of short chain-like or ring-like clusters of hydrogen bonded hydroxyl
groups. Interestingly, this conclusion was reached by several different
routes: the direct reconstruction of the pre-peak as a diffraction
pattern from assumed cluster shapes \cite{EXP_Sarkar_Joarder_Methanol,EXP_Sarkar_Joarder_ethanol},
the thermodynamic route of matching the internal energy with H-bond
associations \cite{Marcus_Water_Octanol}, a hard sphere model for
the spherical cluster aggregates \cite{Frank_Abraham_Lieb_Octanol_Hard_sphere},
and more recently various computer simulation analyses \cite{SimulationLudwigMethanol,AUP_Neat_Alcohols_JCP,AUP_Neat_alcohols_PRE,
SIM_Finci2_linear_alcohols,EXP_Pustai_MethEthProp,SIM_MacCallum_Octanol,
SIM_Siepmann_Octanol,SIM_Mariani_alcohols_pressure,Tomsic_butanol}.
All these descriptions emphasize a causal link between the existence
of clusters formed by the hydroxyl groups and the pre-peak in $I(k)$.
This is directly inspired from the fact that radiation is scattered
off \textit{objects}, usually atoms or molecules, hence specific features
in $I(k)$, such as the pre-peak, should refer to the existence of
corresponding \textit{meta-objects}, such as clusters, aggregates
or self-assembled structures \cite{Nano_reviews_Boldon}. These descriptions
also pose the question as to whether or not it is possible to explain
the features of $I(k)$ in a model free approach.

In the present work, we wish to emphasize that it is not so much the
meta-objects, but the density correlations associated with the atomic
constituents of such objects, which help explain the details in $I(k)$.
In particular, the atom-atom correlation functions reveal differences
in the atomic ordering, depending on the head group atoms (mostly
hydroxyl atoms), and the alkyl tail atoms. These differences are explained
in terms of force field representations by the fact that head group
atoms are charged, while alkyl tail atoms are overall neutral (or
very weakly charged). Hence, the head group atoms tend to cluster
through the Coulomb interaction, mostly into short chain-like aggregates
surrounded by the tail atoms. The corresponding micro-structure affects
the shape of the atom-atom distribution functions, which in turn allow
to infer the local ordering of the various types of atom groups. Such
correlation functions, which depend strongly in the choice of the
force field models, can only be obtained from computer simulations,
enforcing the idea of an unavoidable model based understanding of
the features of $I(k$).

This interpretation is directly supported by the very definition of
the scattering intensity \cite{Debye1,Debye2} as related to the statistical
thermal or ensemble average 
\begin{equation}
I(k)\propto<\sum_{ij}\exp(i\vec{k}\cdot(\vec{r}_{i}-\vec{r}_{j}))>\label{Iraw}
\end{equation}
where the sum runs over all pairs of atomic sites $i,j$. Under this
form, it is not possible to recognize how atomic sites organise themselves
into clusters or aggregates, both inside a molecule and across different
molecules. However, the r.h.s. of the above equation can be written
in terms of atom-atom structure factor, and the Debye formula \cite{Debye1,Debye2}
allows to rewrite Eq.(\ref{Iraw}) as 
\begin{equation}
I(k)=r_{0}^{2}\rho\sum_{ij}f_{i}(k)f_{j}(k)S_{ij}^{(T)}(k)\label{Ik}
\end{equation}
where $S_{ij}^{(T)}(k)$ contains the intra-molecular atom-atom structure
factor $w_{ij}(k)$ (to be discussed in Section \ref{sec:Model}),
and the Fourier transform of the atom-atom intermolecular pair correlation
function $g_{ij}(r)$ 
\begin{equation}
S_{ij}^{(T)}(k)=w_{ij}(k)+\rho\int d\vec{r}\left[g_{ij}(r)-1\right]\exp(i\vec{k}\cdot\vec{r})
\label{STk}
\end{equation}
$\rho=N/V$ is the density (where N is the number of molecules in
the volume V), the $f_{i}(k)$ functions are the form factor of atom
$i$, and $r_{0}=2.8179$ $\cdot10^{-13}$cm is the electronic radius.
Eq.(\ref{Ik}) shows that $I(k)$ is related to the density correlations
of the liquid through the pair correlation functions $g_{ij}(r)$.
Since these quantities describe local atomic and molecular ordering,
any specific feature in $I(k)$ is necessarily related to this local
order, as expressed through the atom-atom structure factors. Therefore,
it is of primary interest to analyse first these structure factors,
in order to better understand the origin of the pre-peak. This is
the route we use herein.

In the present report, we compare the experimental X-ray scattering
intensities $I(k)$ for several mono-ols, with computer simulation results for 
several force fields. Similar comparison have been reported by 
several authors \cite{SimulationBakoNeatMethClusters,
ExpScattBenmoreEthanol,
SIM_Bako_methanol_EXP,ExpScattMatijaMonools,EXP_Pustai_MethEthProp,Tomsic_butanol}, 
with the generic aim of relating the pre-peak to the clustering of
the hydroxyl groups. In the present work, we show that the alkyl tails
also play a crucial role in the interpretation of the pre-peak. This
is due to the fact that the charge ordering of the head groups are
influenced by the ordering of the alkyl tail groups, as witnessed
by the micro-segregation of the charged and uncharged groups observed
in computer simulations.

Charge ordering has been recently studied in the context of room temperature
ionic liquids \cite{Ionic_1,Ionic_2,Ionic_3}, where the polar/apolar
segregation of charged and neutral atomic groups \cite{Triolo_Amphiphile}
provides a contextual link with the present developments. In such
works, the scattering pre-peak is related to the nano-segregation
of the charged and neutral groups, described as polar/apolar in the
corresponding literature. The principal difference with the present
system is the fact that in ionic systems the charges are free to move,
whereas in alcohol molecules they are constrained both to molecular
neutrality and to be attached to the alkyl tails. We come back to
this difference in the Discussion section.

From the remarks above, it becomes very clear that the existence of
a pre-peak in X-ray scattering $I(k)$ cannot be analysed without
detailed molecular simulations, involving in particular approximate
force field models. Therefore, it would be crucial to have a reliable
model capable of reproducing the details of the shape of $I(k)$ for
various mono-ols. The present study reveals that not all models are
able to reproduce the pre-peak behaviour correctly. Hence this comparison
provides a severe selection principle. In particular, the analysis
reveals that the ability of a force field model to reproduce (or not)
the pre-peak in $I(k)$, is less related to a proper account of the
charge and domain ordering, than the ability to properly describe
the incomplete canceling of these two forms of local order.

The remainder of this paper is as follows. In the next section we
describe the X-ray experiments, simulation models and protocols, as
well as the theoretical developments required to analyse the atom-atom
correlation contributions to $I(k)$. The Results section shows detailed
analysis of the X-ray intensities obtained from scattering experiments
and molecular dynamics calculations. A discussion and conclusion sections
close the presentation of this work.

\section{Experimental and model simulation details}

\subsection{Experimental}

The X-ray diffraction (XRD) experiments were performed at the beamlines
BL2 and BL9 of the DELTA synchrotron radiation source using the setup
for wide angle X-ray scattering \cite{krywka2007small} with an incident
X-ray energy of 11\,keV and 13\,keV, respectively. The X-rays were
monochromatized using a multilayer monochromator at BL2 and a Si(311)
double crystal monochromator at BL9 with a beamsize of 0.5x0.5\,mm$^{2}$
and 1x1\,mm$^{2}$ (vxh) at the sample position, respectively. The
scattered photons were detected by a MAR345 image plate detector.
The calibration of the two-dimensional diffraction patterns was performed
with silicon and lanthanum hexaboride references. We measured the
linear mono-ols methanol (purity$\geq$99.9\,$\%$), ethanol ($\geq$99.9\,$\%$),
1-propanol ($\geq$99.9\,$\%$), 1-butanol ($\geq$99.8\,$\%$),
1-pentanol ($\geq$99.8\,$\%$), 1-hexanol ($\geq$99\,$\%$), 1-heptanol
($\geq$99.9\,$\%$), 1-octanol ($\geq$99\,$\%$), 1-nonanol ($\geq$98\,$\%$),
1-decanol ($\geq$99\,$\%$) and 1-undecanol ($\geq$97.5\,$\%$).
All samples were purchased by Sigma Aldrich, except 1-octanol that
was bought from Alfa Aesar, and used without further treatment. The
linear mono-ols were filled into borosilicate capillaries with 3.5\,mm
(BL2) and 2\,mm (BL9) diameter and measured at a temperature of about
293\,K. The XRD images were integrated with the program package Fit2D
\cite{hammersley1996two} and converted to wave-vector transfer $k$
scale. The diffraction pattern were corrected for the scattering contributions
of the capillaries as well as of air and were normalized to the mean
integral of the calculated diffraction patterns in the $k$-range
of 0.2 to 2.3\,$\mathring{A}^{-1}$.

\subsection{Models\label{sec:Model}}

We have used several models for the linear mono-ols, previously studied
in the literature. Namely, we considered the OPLS (Optimized Potentials
for Liquid Simulations) \cite{FF_OPLS_alcohols_1,FF_OPLS_AA_organic_liquids_amines},
TraPPE (Transferable Potentials for Phase Equilibria) \cite{FF_Trappe_Alcohols},
CHARMM (Chemistry at Harward Macromolecular Mechanics) \cite{FF_Charmm1,FF_Charmm2,FF_Charmm3},
and to some extent the GROMOS \cite{FF_Gromos_54a7,FF_Gromos_ATB}
force field models. All models, except for CHARMM are used in their
united atom (UA) versions for the methyl/methylene groups. Indeed,
CHARMM is an all-atom (AA) model by construction. The OPLS AA model
for methanol and ethanol have been tested, in order to have an equivalent
comparison with CHARMM. All force field models are based on the atom-atom
interactions modeled as a Lennard-Jones centre with a partial charge.
Since our interpretation of the pre-peak feature is based on local
charge order, we have listed in Tables 1-5 in the ESI, the typical Lennard-Jones
energy $\epsilon_{i}$ and diameter $\sigma_{i}$, and Coulomb charge
$e_{i}$ for each of the atoms $i$, namely the hydrogen $H$, oxygen
$O$ and first carbon group $C_{1}$, which are charged, and the remaining
carbons numbered as $C_{i}$, for $i=2,n$ where $n$ is the terminal
carbon group. For the CHARMM model, the partial charges of the hydrogen
and the carbon of the carbon groups are explicited. We note that all
models are flexible, hence bond length and dihedral dynamics are considered
consistently with the corresponding parameters in the various force
fields. We will report elsewhere detailed analysis of the differences
between the models concerning the contribution of the local structure
on $I(k)$.

We retain from this section that all UA models attribute a negative
charge for the oxygen atom and positive one for the hydrogen atom,
followed by another positively charged first methylene atom, and remaining
methyl groups are uncharged. From this observation, we can already
predict that the 3 head group atoms will tend to cluster, leaving
the uncharged tail atoms to randomly position around and constrained
by the sole packing effects. If these two groups were separated, the
resulting mixture will readily phase separate into a polar part and
an apolar part. Since these two groups are bound into a single alcohol
molecule, they produce a micro-structure which brings the full demixing
to a local micro-segregation of the polar and apolar group. This is
similar to what is observed ionic-liquids \cite{Triolo_Amphiphile}.
The problem posed by this work is to figure out the link between this
micro-structure and the pre-peak in $I(k)$.

We emphasize again that, unlike the OPLS and TraPPE UA models, the
CHARMM model is an AA model, hence the alkyl tail carbons contain
partial charges over the central atom and the surrounding hydrogen
atoms. Hence, these alkyl tail groups will likewise tend to exhibit
local charge ordering. Although the total charge of the carbon groups
is zero for the remote alkyl groups, there is nevertheless a Coulomb
charge influence in the way these groups will tend to position with
respect to each other and the hydroxyl groups. We will show below
that, although appearing at first more realistic than the AA models,
this model poses some problems such as the absence or suppression
of scattering pre-peak for methanol and ethanol.

\subsection{Simulations}

All simulations were performed with the program package Gromacs \cite{SimulationGromacs}.
We systematically used $N=2048$ number of molecules for all systems,
which corresponds to box sizes ranging from approximately $52$\,$\mathring{A}$,
for methanol to $82$\,$\mathring{A}$\, for 1-nonanol. \.{F}or
1-octanol, simulations of $N=8000$ molecules were performed for testing
purpose, which produced correlation functions not distinguishable
from $N=2048$, indicating that this system size was sufficient even
for longer alkanols.

The systems were simulated in the isobaric-isothermal (constant $NpT$)
ensemble, at the temperature of $T=300$ K and pressure $p=1$ bar.
Those conditions were achieved with the Nose-Hoover thermostat \cite{MD_thermo_Nose,MD_thermo_Hoover}
and Parrinello-Rahman barostat \cite{MD_barostat_Parrinello_Rahman_1,MD_barostat_Parrinello_Rahman_2}.
It is worth emphasizing this procedure allows to test the models under
the same conditions as the experimental ones. As a consequence, the
calculated densities may differ from the experimental ones. We do
not enforce the experimental densities using the isochoric constant
NVT ensemble, since it would creates a bias in the model analysis.
In particular, the correlation functions and the structure factors
differ according to whether these are calculated in the $NpT$ or
$NVT$ ensemble, because of the density differences. This is important
to consider when comparing calculated intensities $I(k)$ obtained
from these different methods.

We followed the same procedure for every simulation. Packmol \cite{MD_Packmol}
was used to obtain the initial configurations of the systems from
the pdb files of each molecule. After energy minimization, the systems
were equilibrated in the $NVT$ and then $NpT$ ensemble, for a total
of 1\,ns each. The following production runs lasted 5\,ns, in order
to sample at least 2000 configurations for calculating the site-site
correlation functions $g_{ij}(r)$. In many cases, several independent
series of such 5\,ns runs were conducted, in order to ensure convergence
of the $g_{ij}(r)$ .

\subsection{Theoretical details}

\subsubsection{The intra-molecular correlations\label{subsec:The-intra-molecular-correlations}}

The evaluation of the total structure factor in Eq.(\ref{STk}) requires
that of the intra-molecular term $w_{ij}(k)$. In previous works,
we had approximated this term by its rigid molecule form \cite{Textbook_Hansen_McDonald}
$w_{ij}(k)=j_{0}(kd_{ij})$, where $j_{0}(x)$ is the zeroth order
spherical Bessel function, and $d_{ij}=|\vec{r}_{i}-\vec{r}_{j}|$
is the inter-atomic distance between atoms $i$ and $j$ on the same
molecule. While this approximation might be acceptable for smaller
molecules, such as ethanol for example (see Fig.1a below), this is
no longer true for longer molecules such as 1-octanol, for example,
for which the flexibility leads to mean inter-atomic distance to be
very different from the rigid molecule values. To account for this,
we have evaluated the intra-molecular correlation functions $w_{ij}(r)$
directly from the configurations obtained from Gromacs, by using the
same standard neighbour histogram method employed for the evaluation
of the inter-molecular correlation $g_{ij}(r)$. The $w_{ij}(k)$
were then evaluated by the same Fast Fourier Technique (FFT) technique
used to obtain the $S_{ij}(k)$ from the $g_{ij}(r)$ \cite{AUP_Neat_Alcohols_JCP}.

In Fig.1 we show for the OPLS force field model the differences between
the rigid molecule version of the $w_{ij}(k)$ (in dashed lines) and
the flexible version (in full lines), for specific atom pairs. Namely,
we focus on 3 alcohols, ethanol, 1-pentanol and 1-octanol, for which
we show $w_{OC_{i}}(k)$ between the oxygen atom $O$ and all the
carbon atoms $C_{i}$ ($i$ ranging from 1 to the terminal one, which
is 2 for ethanol, 5 for 1-pentanol and 8 for 1-octanol), as well as
$w_{C_{1}C_{j}}(k)$, for the correlations between the first carbon
atom and all remaining others. The choice of these atoms allows to
test the flexibility as seen from the oxygen atom to all carbon groups,
as well as that of the carbon atoms between themselves. In all 3 plots,
the gray vertical line marks the pre-peak position in the corresponding
OPLS $I(k)$ (see Fig.4 in the next section).

\begin{figure}[ht]
\centering \includegraphics[height=5cm]{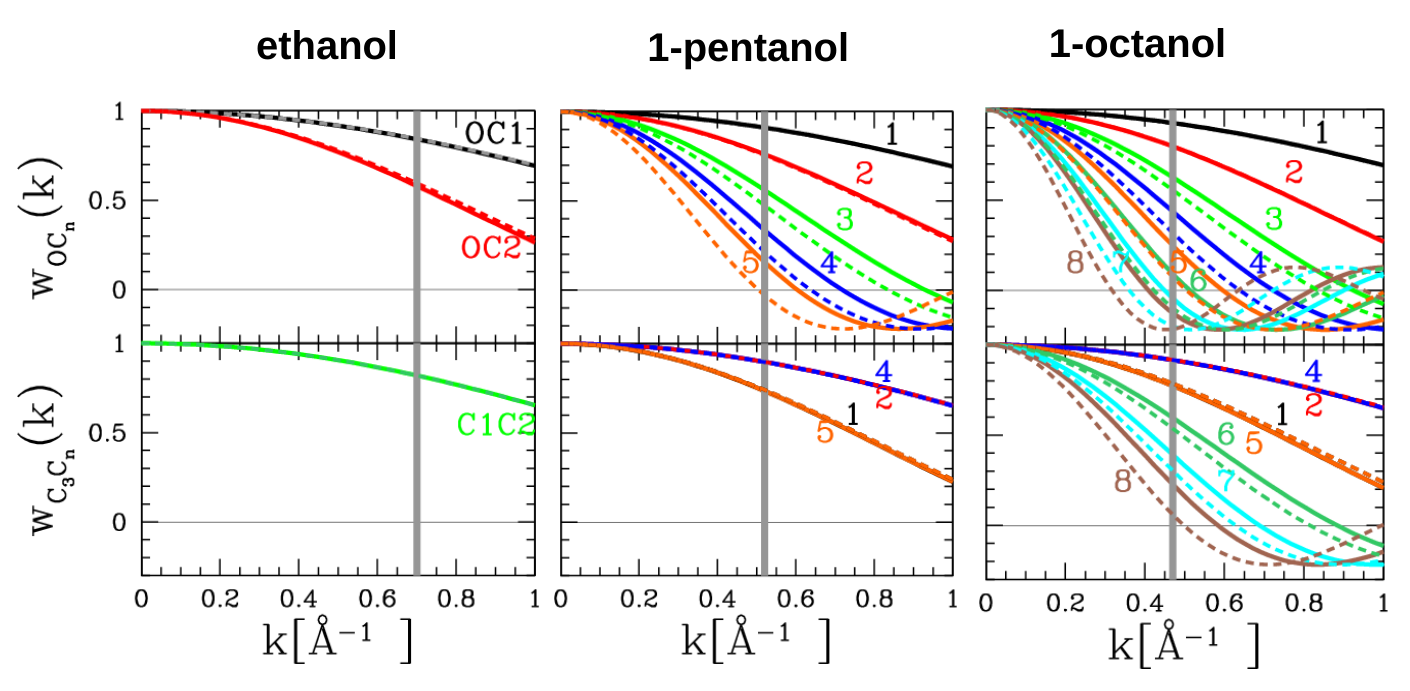} 
\caption{$w_{XC_{1}}(k)$ for ethanol, 1-pentanol and 1-octanol, where X=O
(upper panels) or X=C$_{1}$(lower panels), and $i=1,2,...n$. The
value of $i$ is shown for 1-pentanol and 1-octanol, next to the corresponding
curves. Full curve is with flexibility of alkyl tail and dashed curve
without. See text for more details.}
\label{fig1-wk}
\end{figure}

Fig.1 a) shows that, for the ethanol molecule, the flexibility does
not affect much the intra-molecular correlations, since the full and
dashed lines are mostly superposed, particularly around the pre-peak
region $k_{P}\approx0.7\mathring{A}^{-1}$. For the case of
1-pentanol (middle panel), this is no longer true for the oxygen atom
and the alkyl tail end carbon atoms, starting from $C_{3}$ (upper
panel). This difference is particularly important around the pre-peak
position $k_{P}\approx0.51\mathring{A}^{-1}$. However, the
carbon atoms are not much affected by the flexibility issue. In particular,
we observe that the mean distance between $C_{3}$ and the neighbouring
carbons is nearly the same: $d_{C_{3}C_{1}}\approx d_{C_{3}C_{5}}$
and $d_{C_{3}C_{2}}\approx d_{C_{3}C_{4}}$. These relations remain
true for the case of 1-octanol (right panel). But we also see that
the flexibility issue is aggravated when considering atoms that are
much further apart, as one would indeed expect from a fully flexible
tail. An important feature seen in both panel b) and c) is that the
intra-molecular correlations for the flexible case are above that
of the rigid case. This feature is crucial in determining the sign
of the pre-peak in $I(k)$, as we will see in the results section
\ref{subsec:Correlation-function-analysis}.

\subsubsection{Correlation functions and the computation of $I(k)$ }

The correlation functions obtained in our simulations are quite smooth,
even at the long range part. In some cases, it was necessary to do
additional production runs, in order to ensure that the $g_{ij}(r)$
were well defined in the long range part. The inter-molecular structure
factors were evaluated by FFT techniques from the $g_{ij}(r)$ (the
second term at the right-hand side of Eq.(\ref{STk})), as in all
our previous works \cite{2015_PCCP_benzMH,2016_PCCP_MH_Versus_Clust,AUP_Charge_ordering_prepeak_neat_alc}.
Since the box size $L$ is finite, all our $S_{ij}(k)$ are not reliable
for $k_{\mbox{min}}<2\pi/L$. This is typically $k_{\mbox{min}}\approx0.115\mathring{A}^{-1}$
for methanol, and $k_{\mbox{min}}\approx0.072\mathring{A}^{-1}$
for 1-nonanol . These values are much smaller than the respective
pre-peak positions $k\approx1\,$$\mathring{A}$$^{-1}$ and $k_{P}\approx0.4\,\mathring{A}^{-1}$,
ensuring that the box size does not affect the physical features we
discuss herein.

The numerical evaluation of $I(k)$ in experimental units of cm$^{-1}$
has been described in \cite{AupPropylamine2}. We remind that this
requires to replace the frontal $r_{0}^{2}\rho$ term of $I(k)$ in
Eq.(1) by the following pre-factor $r_{0}^{2}(N/L^{3})/100$, $N$
is the number of molecules in the simulation box of size $L$(expressed
in meters).

\section{Results\label{sec:Results}}

\subsection{Experiments}

The experimental $I(k)$ are shown in Fig.2 for the measurements performed
at BL2 of DELTA and are in reasonable agreement with the data published
by Vahvaselk\"a et al. \cite{ExpScattFinnsMonools} and Tom\u{s}i\v{c} et
al. \cite{ExpScattMatijaMonools}\cite{Tomsic_butanol} (see Fig.1
in SI).

\begin{figure}[ht]
\centering \includegraphics[height=6cm]{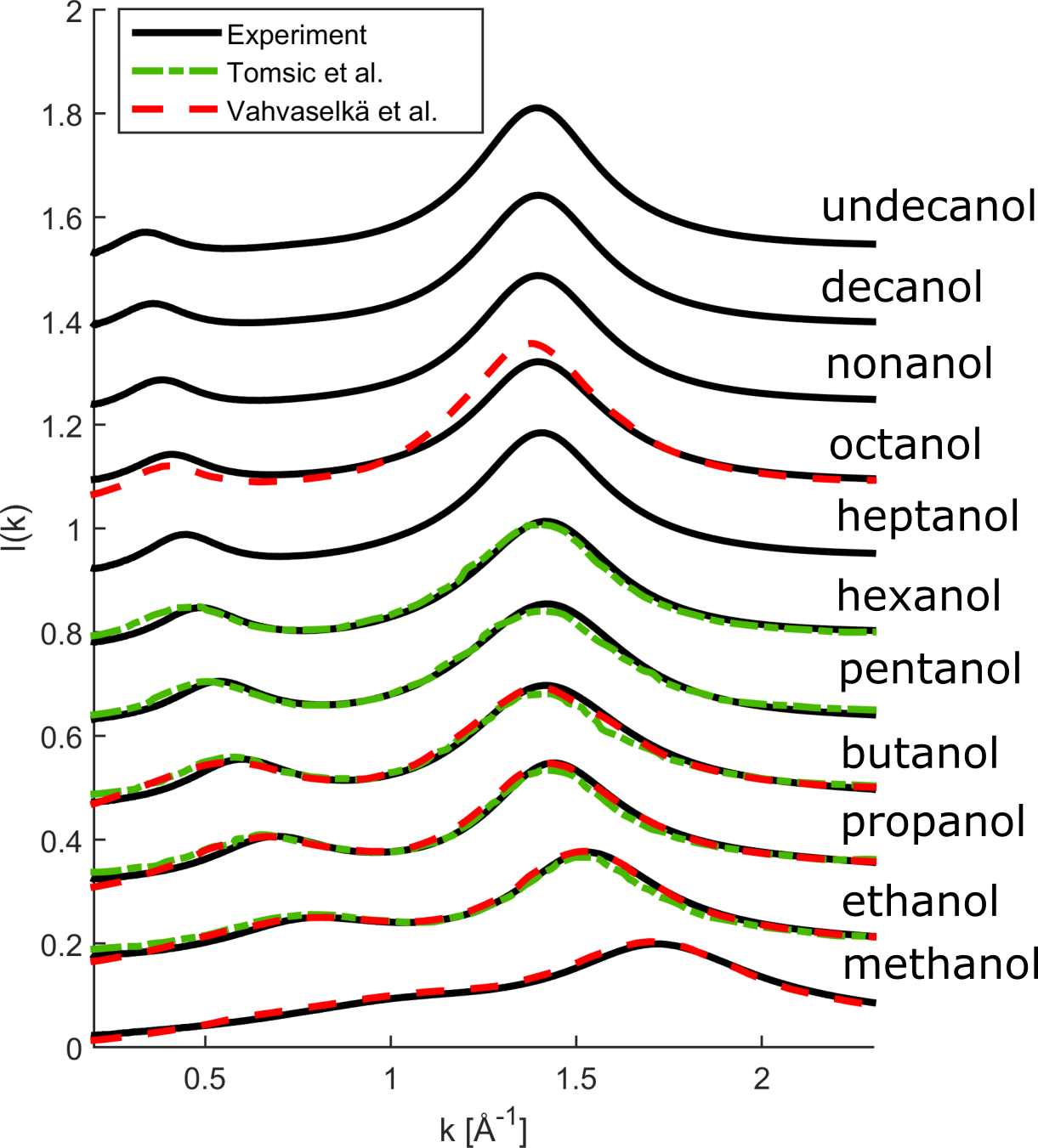}
\caption{Experimental X-ray scattered intensities for alcohols from methanol
to 1-nonanol. The inset shows the pre-peaks, as extracted from a procedure
explained in Section 3.3.}
\label{fig2-Ik}
\end{figure}

Despite the different experimental setup used at BL9 of DELTA, the
data taken at both experiments resemble each other. Complementary
measurements were performed on a D8 laboratory diffractometer with
less statistical accuracy but are also consistent with the ones presented
here. The patterns exhibit two distinct peaks, the main diffraction
peak and the well known pre-peak feature in mono-ols \cite{ExpScattMaginiMethanol,ExpScattNartenEthMeth,EXP_Sarkar_Joarder_Methanol,EXP_Sarkar_Joarder_ethanol,ExpScattFinnsMonools,ExpScattJoarderAlcohols,ExpScattMatijaMonools,EXP_Sarkar_1propanol,EXP_Joarder_Methanol_temp,EXP_Pustai_MethEthProp,EXP_Silleren_Propanol}.
The latter can be observed in the $k$-range $0.3\,\mathring{A}^{-1}$<k<1\,$\mathring{A}^{-1}$
and is well separated from the main diffraction maximum ($1.4\,\mathring{A}^{-1}$<k<1.75\,$\mathring{A}^{-1}$).
Interestingly, the pre-peak is a simple shoulder for methanol. The
main peak position and amplitude varies considerably from methanol
to 1-propanol, above which it stabilises in position while increasing
in amplitude. As for the pre-peak, two important features are observed.
The first concerns the systematic decrease of the pre-peak position
with increasing alcohol length. A second feature is the fact that
the pre-peak amplitude increases with $n$ for small alkanols, until
a maximum for 1-butanol, after which it decreases systematically.
This behaviour suggests a crossover phenomena, which we will study
in the following sections. Both observations are line with a previous
similar report of these trends \cite{Tomsic_Brij35_Water_Alcohols},
and require an explanation.

\subsection{Simulations}

The results for $I(k$) from computer simulations, and for various
models are shown in Fig.3, and are compared with the experimental
ones as shown in Fig.2. The results are presented for the OPLS model
in blue, for the TraPPE model in red, for the CHARMM model in green
and the GROMOS model in gold, while the experiment is plotted as black
dashed lines. The gray area represents the small-$k$ region for which
the finite size errors from the simulations affect the estimations
of the calculated $I(k)$.

\begin{figure}[ht]
\centering \includegraphics[height=9cm]{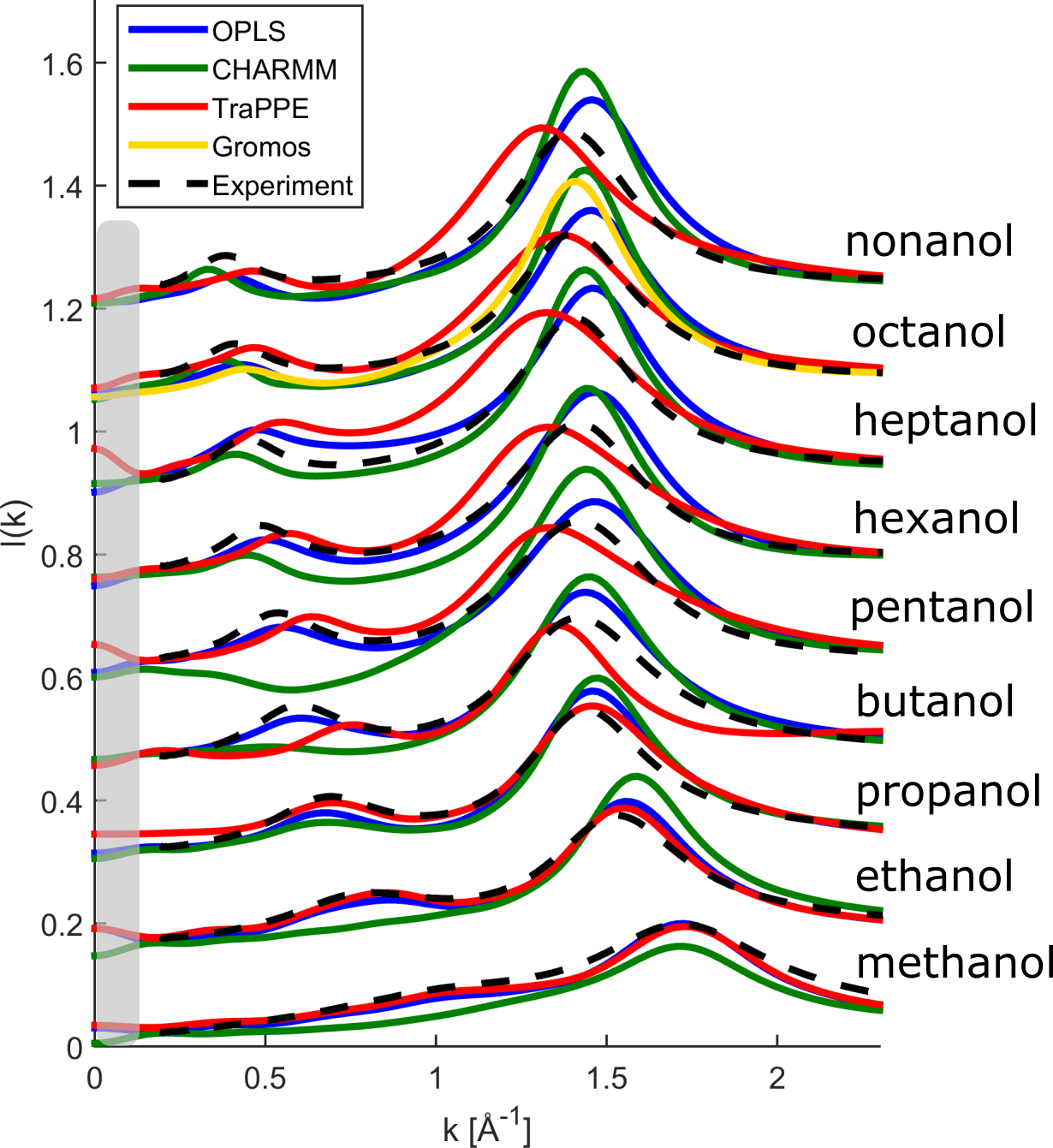} 
\caption{$I(k)$ calculated with the OPLS (blue), TraPPE (red), CHARMM (green),
GROMOS (gold) model from methanol to 1-nonanol compared to the experimental
diffraction patterns (black dashed lines).}
\label{fig3-IkSim}

\end{figure}

It is immediately seen that, for all models, the overall shapes of
the calculated $I(k)$ for various mono-ols are in agreement with
the experimental ones, including in particular the main peak in the
range $1.3\,\mathring{A}^{-1}<k_{M}<1.7\,\mathring{A}^{-1}$, and
the pre-peak in the range $0.3\,\mathring{A}^{-1}<k_{P}<1.0\,\mathring{A}^{-1}$.
The OPLS model seems more consistent than the two others, in the sense
that it reproduces both main peak and pre-peak features better than
the other models. It generally overestimates main peak height and
underestimates pre-peak height. The TraPPE model does not provide
a good description of the main peak for longer mono-ols (amplitude
and position), but produces a pre-peak consistent with experiments.
The position of the pre-peak is in remarkable agreement with the experiment
up to 1-propanol, then deviates significantly but approaches again
the experimental pre-peak position for the largest mono-ols. CHARMM
model is the most inconsistent one, since it is unable to produce
the pre-peak for lower alkanol (except for 1-propanol - but see below).
Interestingly, this model predicts an unphysical negative pre-peak
(anti-peak) for 1-pentanol. It predicts consistent pre-peaks for longer
chains. 

The good agreement at the main peak indicates that the force field
model is able to describe short range spatial correlations relatively
well. Most force fields are designed to achieve this to some extent.
Indeed, most force fields give a good account of thermodynamic properties
such as the enthalpy. This quantity is related to the elementary integrals
$\int d\vec{r}g_{ij}(r)v_{ij}(r)$, where $v_{ij}(r)$ is the pair
interaction modeled by the force field, for a given pair of atomic
sites $i$ and $j$. Such interactions have a spatial range limited
to first or second neighbours, even in presence of long range interactions,
because of the Coulomb screening phenomena in dense liquids. Therefore,
a good approximation of the enthalpy indicates that both the interaction
$v_{ij}(r)$ and the resulting pair correlation $g_{ij}(r)$ must
be well reproduced by the force field model.

A good agreement for the pre-peak is a more demanding feature, indicating
that the force field is able to capture cluster/domain formation accurately.
This information is hidden in the long range part of the correlations,
and is more sensitive to the microscopic details.

A systematic analysis of both main peak and pre-peak features is provided
in the next sub-section.

\subsection{Main peak and pre-peak analysis}

For a detailed analysis we determined the main-peak using a combination
of a Pearson VII function and a linear slope. Then in a second step
we fitted the pre-peak likewise in the corresponding pre-peak region
after subtraction of the main-peak fit and revealed their peak intensity,
peak position and FWHM (Full Width at Half Maximum). The corresponding
error bars were estimated analyzing the measurements carried out at
the various experimental end-stations and are dominated by the background
treatment. This procedure was likewise performed for the computed
$I(k)$ for proper comparison between experiment and theory. 
 However, using this procedure it was not possibly
to extract a pre-peak for all the computed $I(k)$, i.e. for methanol,
ethanol, 1-butanol and 1-pentanol obtained from CHARMM force field.
The error bars of this analysis were obtained by variation of the
corresponding fitting ranges. The results of the evaluation are shown
in Fig. 4 for the main peak and Fig. 5 for the pre-peak.

\begin{figure}[ht]
\centering \includegraphics[height=5cm]{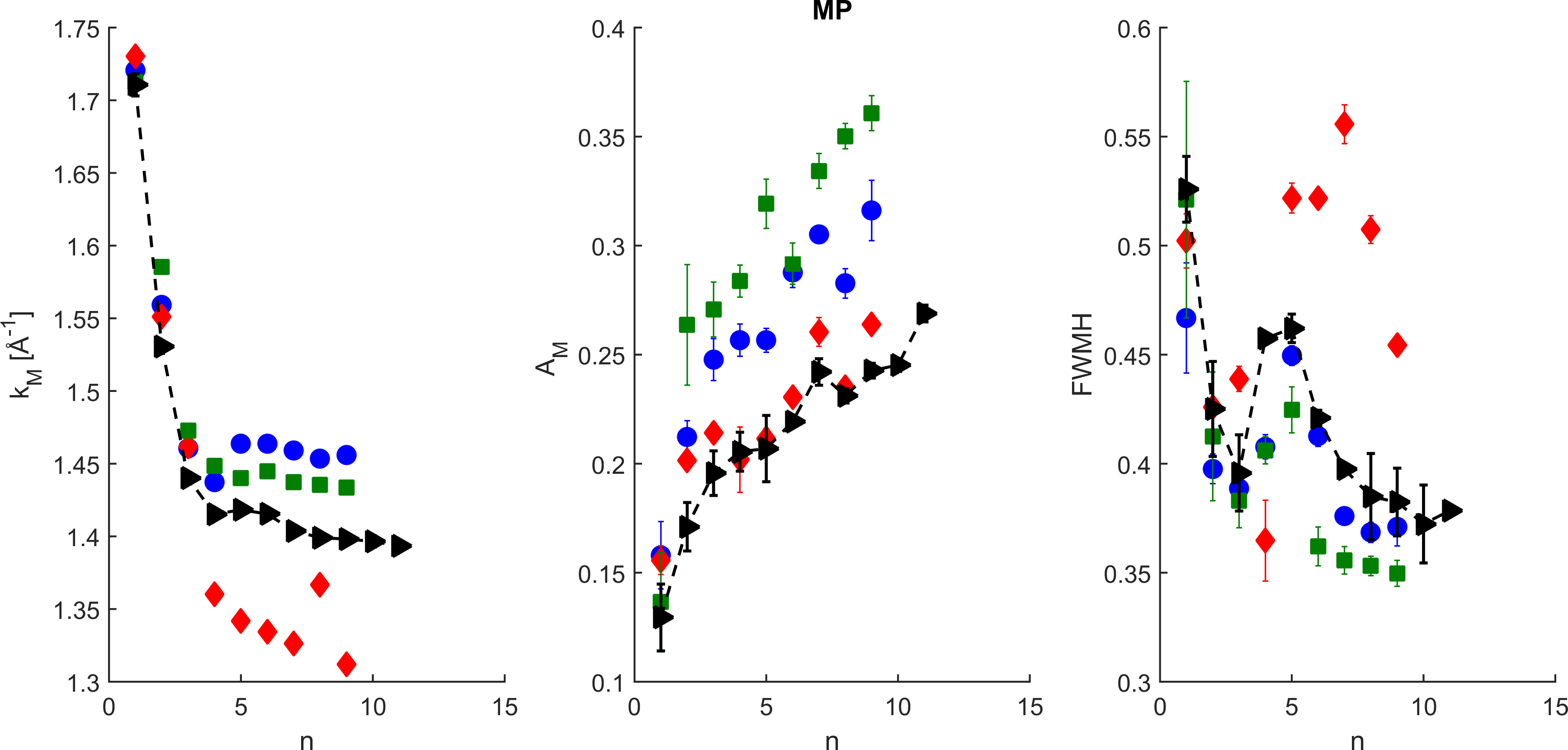} 
\caption{Main peak analysis of $I(k)$ from the experiment (black triangles),
OPLS model (blue dots), CHARMM model (green squares) and TraPPE model
(red diamonds). Peak position $k_{M}$ (left), peak amplitude $A_{M}$
(center) and FWHM (right).}
\label{fig4-MP} 
\end{figure}

Fig.4 shows that the main peak amplitude $A_{M}$ increases with alcohol
chain length showing an almost linear behavior for longer mono-ols
starting with 1-propanol, while the position $k_{M}$ strongly decreases
and then changes almost linearly with chain length. Surprisingly,
the FWHM exhibits a clear non-linear dependency. This can be assigned
to the proximity of the position of both pre-peak and main peak up
to 1-butanol and the corresponding crosstalk in the fit. For larger
chain lengths the difference in peak positions strongly increases
providing a more independent fit of both features. However, this behavior
can be observed consistently in simulations (OPLS and CHARMM) and
experiment.

These trends in the main peak can be rationalised as follows. The
position $k_{M}$ of the peak is related to the mean atomic diameter
$<\sigma>$ for a given alcohol, according to $k_{M}\approx2\pi/<\sigma>$.
For a dense liquid, this is equivalent to $<\sigma>$ being related
to the average particle-particle contact. For small alcohols, this
is dominated by the small diameter of hydrogen, hence a larger $k_{M}$.
But for the longer ones it would be dominated by the diameter of the
methylene group, hence a smaller $k_{M}.$ Since, for longer alcohols
the methylene groups dominates the averaging, it explains why $k_{M}$
tends to saturate at the lower value of $1.4\mathring{A}^{-1}$, which corresponds
to a $\sigma\approx 4.5\mathring{A}^{-1} $, which is close to 
 $\sigma_{CH_2}\approx4.3\mathring{A}$. This is consistent for
OPLS and CHARMM, but not TraPPE, which indicates too a large methylene
diameter about $4.65\mathring{A}$. The amplitude $A_{M}$ is governed, according
to Eq.(\ref{Ik}), by the main peak amplitude of the various structure
factors $S_{ij}(k)$. 
Although the number of methylene-methylene
correlations increases with $n^2$, 
the density term $\rho=N/V$ in Eq.(\ref{Ik}) moderates this trend,
since simulations indicate that, for a fixed number of molecules $N=2048$,
the box size $V=L^{3}$ increases with $n$.

Overall, the OPLS model appears to follow the overall experimental
trends better than the CHARMM and TraPPE models.

Fig.5 shows a similar analysis for the pre-peak.

\begin{figure}[h!]
\centering
\includegraphics[height=5cm]{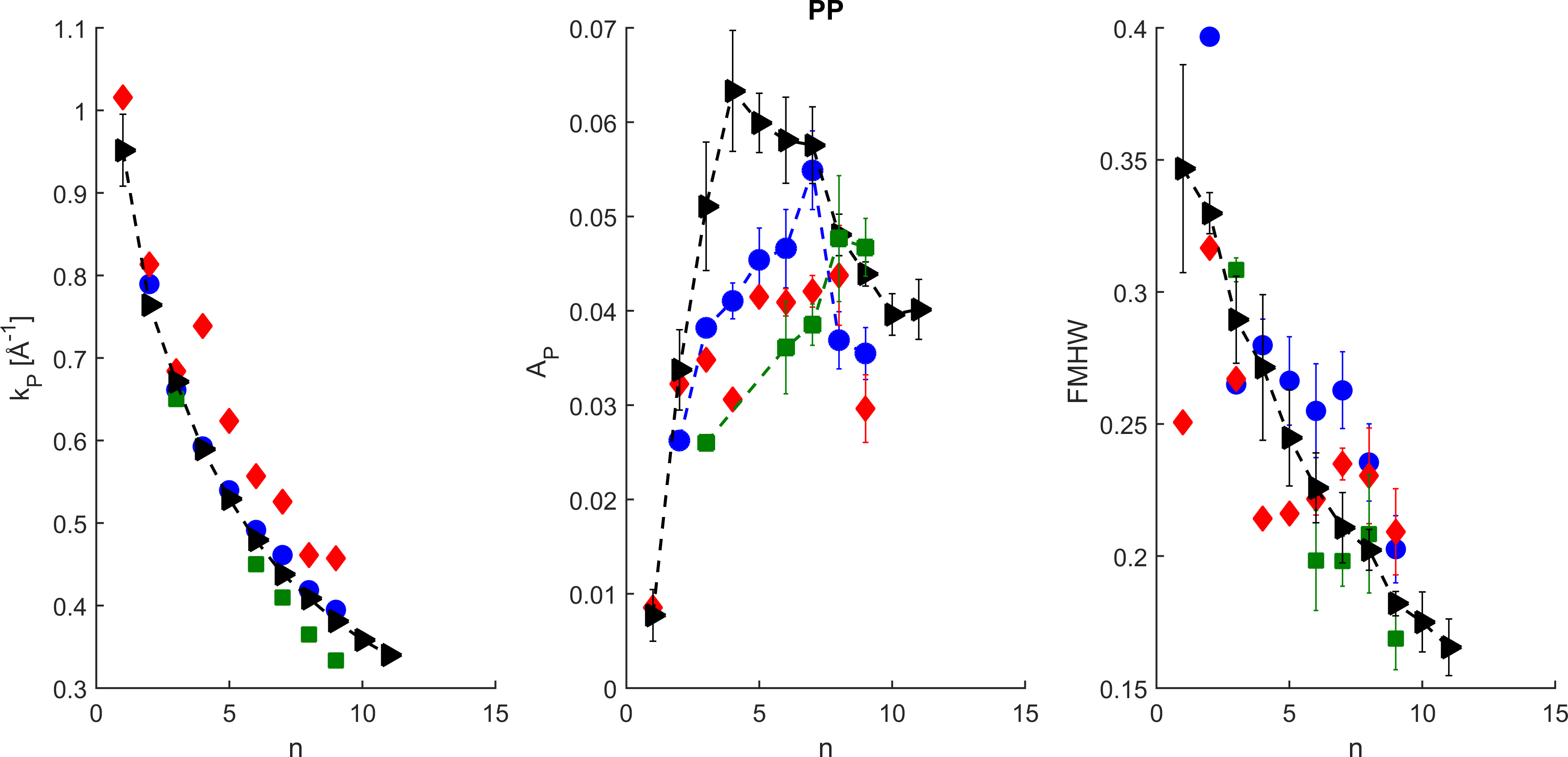} 
\caption{Pre-peak analysis of $I(k)$ from the experiment (black triangles),
OPLS model (blue dots), CHARMM model (green squares) and TraPPE model
(red diamonds). Peak position $k_{P}$ (left), peak amplitude $A_{P}$
(center) and FWHM (right).}

\label{fig5-PP} 
\end{figure}

The pre-peak position $k_{M}$ is seen to decrease monotonously with
$n$. This trend can be understood the following way. $k_{M}$ is
related to the size of the meta-object, but which? Is it the mean
length of the hydroxyl head group linear aggregate? 
The cluster analysis (see Fig.6 below) shows that the cluster size is either dominated 
by pentamers for all mono-ols (OPLS, TraPPE) or changes very little (CHARMM), 
hence in contradiction with the trend observed in Fig. 4. 
It seems more reasonnable to associate the
meta-object to the ensemble of the chain aggregate surrounded by the
alkyl chain, whose mean size $<d>$ would increase with $n$, in agreement
with the observed behaviour for $k_{P}\approx2\pi/<d>$. The FWHM
of the pre-peak decreases continuously from methanol to 1-undecanol.
As for the amplitude $A_{P},$ the experimental data show that it
increases up to 1-butanol , saturates, and then decreases starting
with 1-octanol. A similar observation for the increase of pre-peak
was made by Tom\u{s}i\v{c} et al. \cite{Tomsic_butanol}, in very good
agreement with our findings. The decrease of the amplitude $A_{P}$
is more complicated to rationalize. It is related to the importance
of the alkyl tail contributions with increasing $n$, and will be
discussed later in the light of the simulations.

Overall by comparison between experiment and theory, the OPLS-UA model
provides the most prominent agreement with the experimental findings.

\subsection{Snapshots}

Snapshots of typical alcohols are shown in Fig.6, and for
various models. Visual inspection confirms mostly the existence of short chain-like
aggregates for all models and alcohols. 

\begin{figure}[ht]
\includegraphics[height=11cm]{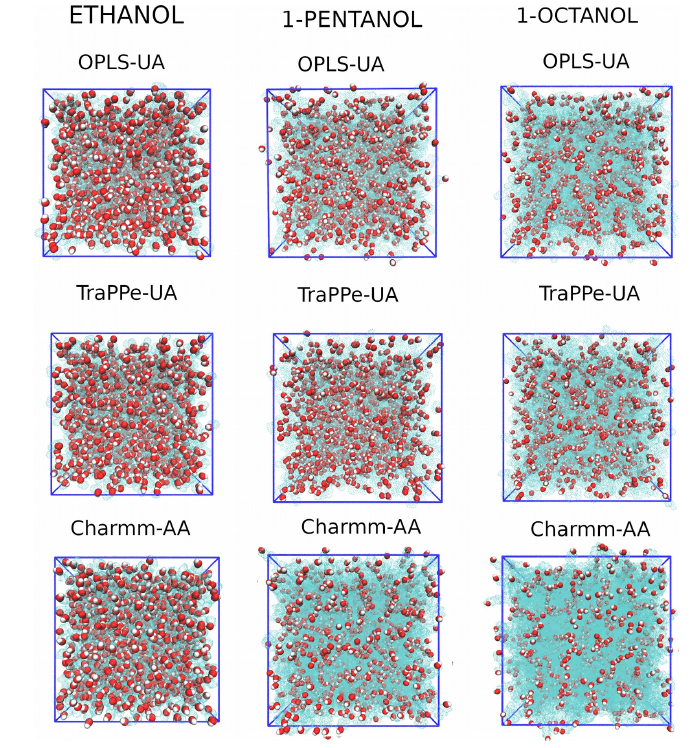} \label{fig6-snap}
\caption{Snapshots of 3 different models (OPLS,TraPPE and CHARMM) for 3 different
alcohols (ethanol, 1-pentanol and 1-octanol)}
\end{figure}

These snapshots are designed to highlight the hydroxyl groups, with
oxygen in red and hydrogen in white, while the alkyl tails are shown
as semi-transparent cyan. While the density of the H-bonded chains
decreases with increasing alkyl chain length, which is obvious since
the number of methyl groups increases, it clearly appears that the
H-bonded chains seem longer for longer mono-ols. This feature can
be rationalized by the fact that charged groups can cluster more easily
in large groups when the alkyl chain is longer. The visual inspection
of the CHARMM model is complicated by the fact that this is an AA
model, hence the alkyl tails are fatter because of the prominent hydrogen
atoms. This makes the cyan background somewhat denser and hinders
the clear vision of hydroxyl chains, and gives a false impression
of a lesser density of the hydroxyl groups. Despite this drawback,
the somewhat lesser clustering tendencies of this model are 
visually apparent.
A more detailed investigation
reveals that all sorts of geometry and size of clusters exists, such as 
linear chains, branched chains, loops, branched loops, lassos. In that, 
clustering shape in all mono-ols is very similar to what we reported earlier for 
methanol \cite{AUP_Neat_Alcohols_JCP}, with the exception that larger clusters 
more prominently appear for longer mono-ols.

\subsection{Cluster analysis}

Fig.7 shows the clustering probabilities of the oxygen atoms for selected
mono-ols and models. The methodology used is the same as in our previous
works \cite{AUP_Neat_Alcohols_JCP,Lara_Ethanol_MolSim}.

\begin{figure}[ht]
\centering \includegraphics[height=6cm]{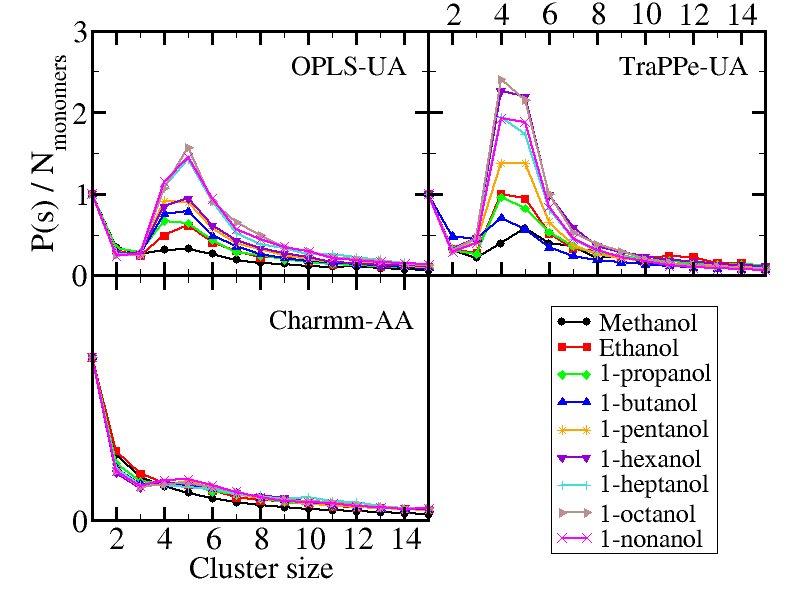} \label{fig7-clust}
\caption{Probabilities of oxygen clusters versus the cluster size, for alcohols
from methanol to 1-nonanol, and for 3 different models (OPLS, TraPPE
and CHARMM)}
\end{figure}

Both OPLS and TraPPE model show a systematic trend for higher clustering
probability with increasing alcohol length, both with an average chain
length around 5. This is somewhat consistent with what is observed
in the snapshots of Fig.6. In particular, for longer alcohols, the
probability of observing pentamers of hydroxyl groups exceeds that
of free monomers. But, the most striking and unexpected feature is
the near absence of a cluster peak for the CHARMM model. 
For this model, the monomer probability is the highest, 
suggesting that the number of pentamer hydroxyl groups is small.
However, the total number of hydrogen-bonded molecules is still 
much larger than the number of monomers. In that, the CHARMM model
is not different than other model, as far as the global clustering
property is concerned.
It confirms what the visual inspection of the snapshots in Fig.6 suggests. 
This problem
may be related to the absence of a pre-peak for this model for many
alcohols, as shown in Fig.3. This marked difference with the other
models requires some clarifications, which the analysis of the correlation
functions is likely to provide.

\subsection{Correlation function analysis\label{subsec:Correlation-function-analysis}}

Since atom-atom structure factors $S_{ij}(k)$ appear in the definition
of $I(k)$ through Eq.(\ref{Ik}), in this entire section we will
analyse how the specific micro-structure of various types of the alcohols
we have studied herein, contribute to the $S_{ij}(k).$ In the remainder,
we choose to graphically represent the inter-molecular part of the
structure factor with a convention similar to a single atom liquid
\cite{Textbook_Hansen_McDonald}, namely:

\begin{equation}
S_{ij}(k)=1+\rho\int d\vec{r}\left[g_{ij}(r)-1\right]\exp(i\vec{k}\cdot\vec{r})\label{Sk_ours}
\end{equation}
This way, aside the common additive term 1, we will be essentially
comparing the Fourier transforms of the atom-atom inter-molecular
correlation functions $g_{ij}(r)$.

\subsubsection{Charge ordering and domain ordering}

From our previous work \cite{2016_JCP_ethMeth,AupPAC,AUP_Charge_ordering_prepeak_neat_alc},
the key feature in the hydroxyl group chaining is the charge ordering
between the negatively charged oxygen and positively charged hydrogen,
leading to a characteristic linear chaining of the type $+-+-+-...$.
As a consequence, the associated atoms having many of their like neighbours
aligned, they have less second and third neighbours of their kind.
This leads to two specific features \cite{AUP_Charge_ordering_prepeak_neat_alc}
in the corresponding inter-molecular $g_{ii}(r)$ for the correlations
between like atoms $i$: a high first peak, witnessing the strong
association, followed by depleted correlations, due to lesser neighbours
of like atoms. This feature is illustrated in Fig.8 for the oxygen-oxygen
correlation functions $g_{OO}(r)$ of the OPLS model for selected
3 mono-ols (ethanol(green), 1-pentanol (blue) and 1-octanol (black)).

\begin{figure}[ht]
\centering \includegraphics[height=6cm]{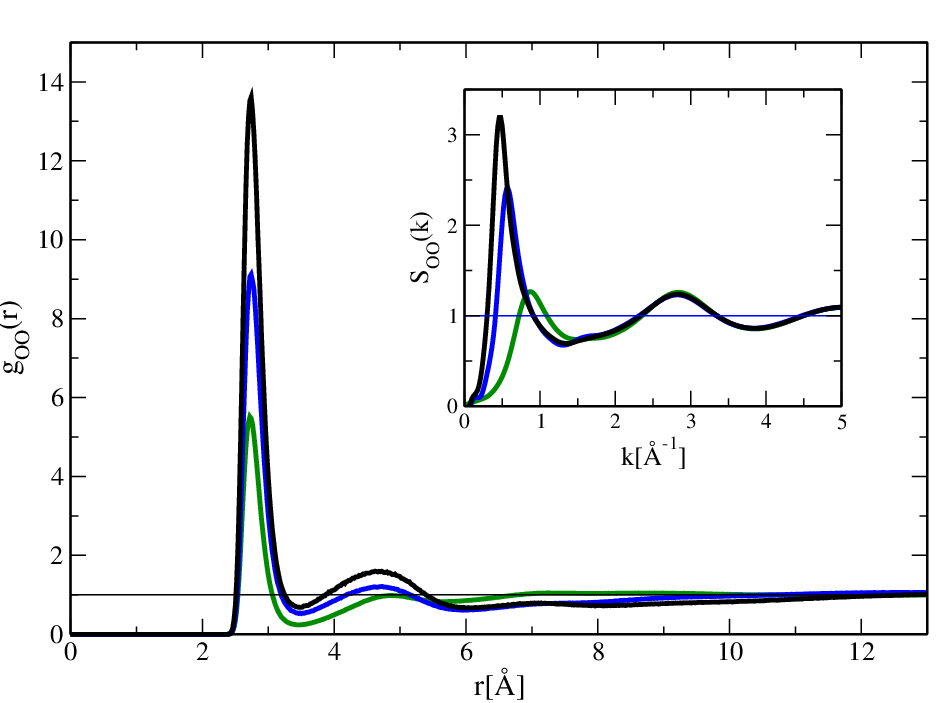} \label{fig8-depl}
\caption{Oxygen-oxygen correlation functions $g_{OO}(r)$ of the OPLS model
for selected mono-ols (ethanol (green), 1-pentanol (blue) and 1-octanol
(black)), illustrating the features which give rise to the structure
factor pre-peak (see text). The inset shows the corresponding pre-peak
in $S_{OO}(k)$.}
\end{figure}

It is this double feature which leads to the pre-peak feature in the
corresponding $S_{OO}(k)$, as was demonstrated in Ref. \cite{AUP_Charge_ordering_prepeak_neat_alc}.
The essence of the demonstration is the following and holds for any
pair of associated atoms $a$: the high first peak of $g_{aa}(r)$
contributes to a wide positive peak at $k=0$ for $S_{aa}(k)$, while
the depletion part of $g_{aa}(r)$ contributes to a narrow negative
peak at $k=0$. The total contribution is a positive pre-peak in $S_{aa}(k)$.
This is illustrated in the inset of Fig.8 for $S_{OO}(k)$.

Another feature which is fundamental to understand the pre-peak in
$I(k)$, is the existence of anti-correlations between the charged
and uncharged groups \cite{AupDomainOrdering,AupPropylamine2}. Indeed,
since charged groups prefer to associate together, the uncharged groups
tend to occupy the remaining empty space, with the permanent constraint
that these groups are attached to the full molecules. The spatial
alternation of charged and uncharged atomic groups is illustrated
in Fig.9 for the OPLS model and 3 selected alkanols. 

\begin{figure}[ht]
\centering \includegraphics[height=5.5cm]{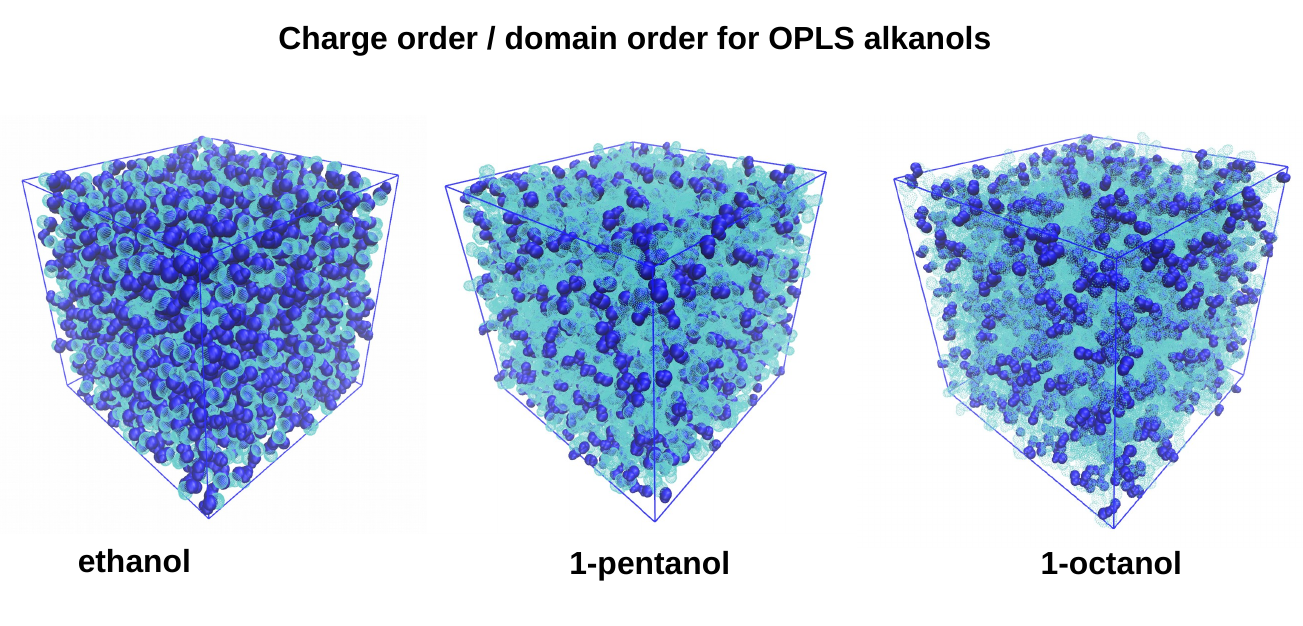} \label{fig9-snapDO}
\caption{Snapshots for 3 alcohols (ethanol, 1-pentanol and 1-octanol) for the
OPLS model, illustrating the nano-segregation between the charged
atoms (dark blue) and the uncharged ones (pale blue, shown in semi-transparent).}
\end{figure}

The charged groups (oxygen, hydrogen and first methylene atom) are
shown in blue, while the remaining methylene/methyl atoms are shown
in semi-transparent cyan. The charged groups form compact semi-linear
blue clusters, separated from the surrounding cyan neutral atoms.

This alternation of groups/domains, leads to long range anti-correlations
between them. This is illustrated in Fig.10 for oxygen-oxygen $g_{OO}(r)$
and oxygen- terminal methyl group $g_{OC_{n}}(r)$ correlation functions,
for the OPLS models of the 3 alkanols selected in the previous snapshots
in Fig.9 .

The long range phase opposition domain oscillations can be observed
in all 3 cases, but these are broader for the longer mono-ols. The
width of the oscillations period increases with the alkanol chain
length. The inset shows the first peak details. The influence of the
Coulomb induced large first peak is clearly seen in the $g_{OO}(r)$,
which is a direct consequence of the H-bonding correlations between
the oxygen atoms.

\begin{figure}[ht]
\centering \includegraphics[height=5cm]{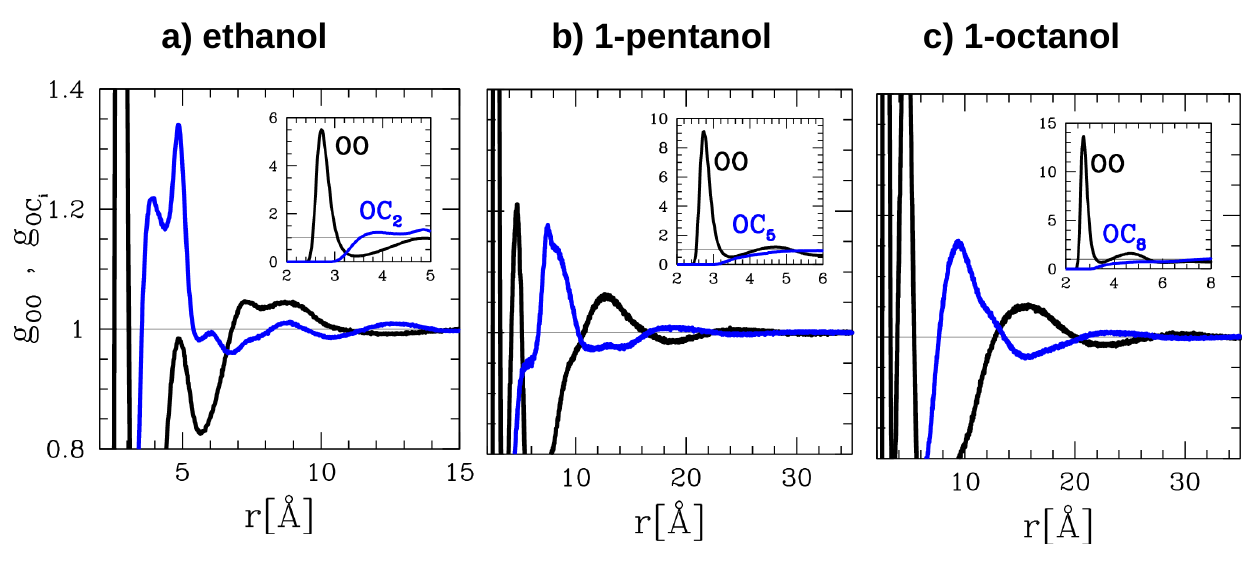} \label{fig10-DO}
\caption{Correlation functions for OPLS model $g_{OO}(r)$(blue) and $g_{OC_{n}}(r)$(black)
illustrating the domain ordering between charged head group atoms
(here O) and tail atoms (here the last carbon C$_{\ensuremath{n}}$),
and for 3 different alcohols (ethanol, 1-pentanol and 1-octanol).}
\end{figure}

These anti-correlations give raise to negative pre-peaks (or anti-peaks)
in the structure factors, as illustrated in Fig.11. The anti-peaks
are easy to understand: since the correlations $g_{OC_{n}}(r)$ are
in opposing phase with those of the $g_{OO}(r)$, their contribution
to the Fourier transform in Eq.(\ref{Sk_ours}) has necessarily the
opposite sign, hence contribute in the opposite direction than that
of the $S_{OO}(k)$ pre-peak.

\begin{figure}[ht]
\centering \includegraphics[height=7cm]{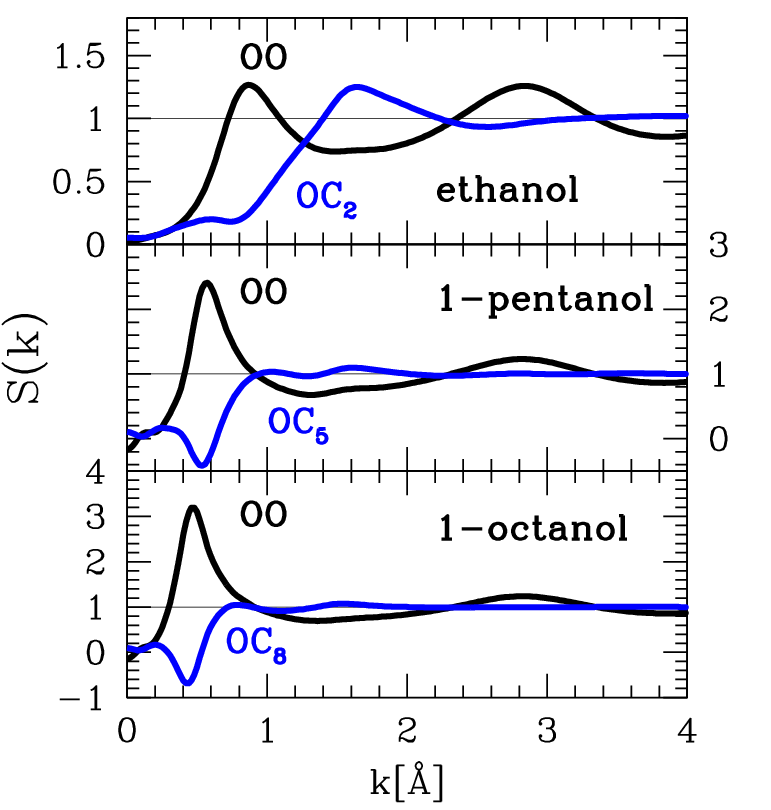} \label{fig11-SkDO}
\caption{Structure factors $S_{OO}(k)$ and $S_{OC_{n}}(k)$ corresponding
to the correlation functions shown in Fig.10.}
\end{figure}

While the positive pre-peaks are quite prominent, and greater than
the main peaks, the anti-peaks are often negative. It is interesting
to note that for the smaller mono-ols, the anti-peak is merely a small
dip (top panel in Fig.11). This is a direct consequence of the alkyl
tails being very short, and being only weakly correlated with the
charged head groups. However, for longer alcohols, these tails are
strongly anti-correlated with the head groups. This anti-correlation
reflects the nano-segregation observed in the snapshots Fig.6, between
the aggregated hydroxyl groups and the alkyl tails. Indeed, if the
two types of groups, charged hydroxyl and neutral tail, were detached,
one would observe a macroscopic phase transition between them. Since
they are tied into the same molecule, one observes a depleted correlation
between the two, as that seen in $g_{OC_{n}}(r)$ in Fig.10, which
is at the origin of the anti-correlation.

Now, we are in position to understand how these correlations and anti-correlations
play a capital role in the pre-peak of $I(k)$, as a result of cancellations
in Eq.(\ref{Iraw}). The lower panels of Fig.12 show all the atom-atom
structure factors $S_{ij}(k)$ (as defined in Eq.(\ref{Sk_ours}),
that is with the extra 1 added for presentation purpose) for the same
3 OPLS-UA alkanols, highlighting the like and cross correlations in
the pre-peak region. The upper panels show the $I(k)$ (in black),
together with the like correlation contributions (in blue) and cross
correlation contributions (in red).

\begin{figure}[ht]
\centering \includegraphics[height=6cm]{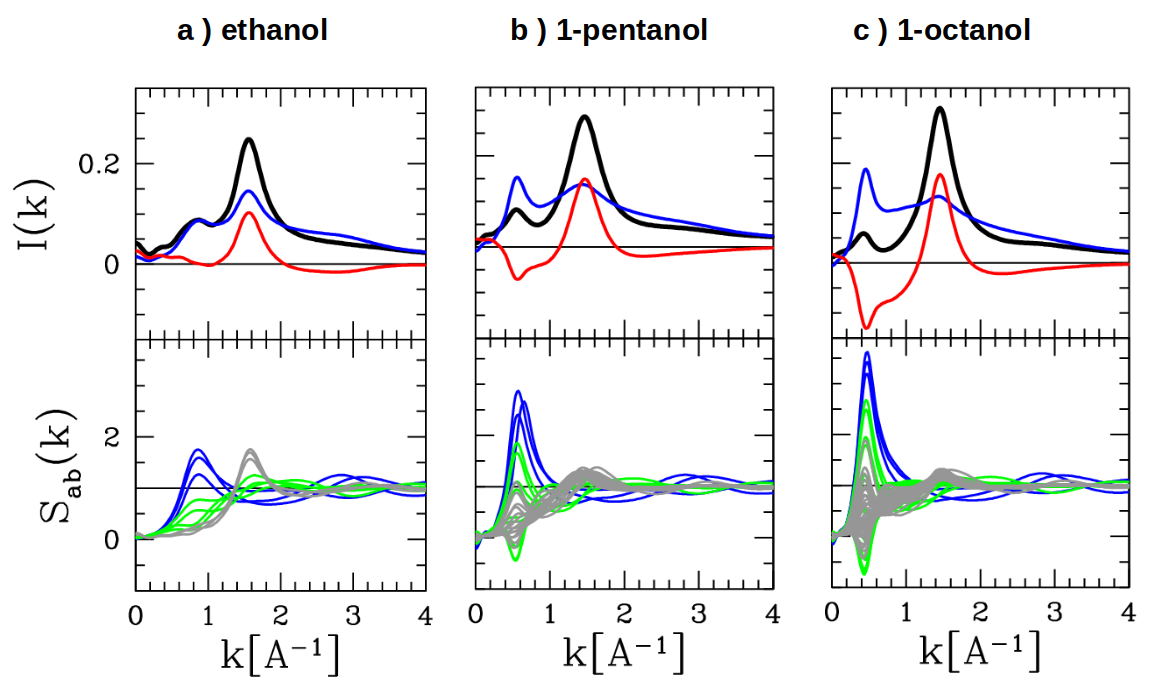}
\label{fig12-IkSk} \caption{Scattered intensities $I(k)$ (upper panels) and all corresponding
atom-atom structure factors $S_{ij}(k)$ for the OPLS model of 3 alcohols
(ethanol, 1-pentanol and 1-octanol). The upper panels equally show
the like atom contributions to $I(k)$ in blue, and cross atom contributions
in red. The lower panels show hydroxyl atom contributions in blue,
alkyl tail carbon atom contributions in gray, and cross contributions
in green (see text).}
\end{figure}

The structure factors in the lower panel are divided into those of
the hydroxyl groups (in blue), the methyl groups (in gray) and the
cross correlation between the two (in green). The specificity of the
various atoms are \emph{intentionally} ignored, since we want to highlight
the influence of head groups versus the alkyl tail, which is at the
origin of the local segregation in Fig.6 and Fig.9. The $I(k)$ in
Eq.(\ref{Ik}) can be rewritten as 
\begin{equation}
I(k)=I_{\mbox{like}}(k)+I_{\mbox{cross}}(k)\label{IsIc}
\end{equation}
with 
\[
I_{\mbox{like}}(k)=r_{0}^{2}\rho\sum_{i}f_{i}^{2}(k)S_{ii}^{(T)}(k)
\]
\[
I_{\mbox{cross}}(k)=r_{0}^{2}\rho\sum_{i\neq j}f_{i}(k)f_{j}(k)S_{ij}^{(T)}(k)
\]
where we have separated the like atom contributions $I_{\mbox{like}}(k)$
from those from the different atoms $I_{\mbox{cross}}(k)$. From the
lower panels of Fig.12, it is clear that $I_{\mbox{cross}}(k)$ is
likely to contain negative anti-peak contributions in the case of
some alcohols, such as 1-pentanol and 1-octanol, for example. The
upper panels of Fig.12 show in black lines the total $I(k)$ built
from all total structure factors from Eqs.(\ref{Ik},\ref{STk}),
as well as the contributions to $I(k)$ coming from like atoms and
cross atomic contributions. Comparing the structure factors in the
lower panel, to the $I(k)$ in the upper one, it is clear how the
like contributions tend to contribute positively to the pre-peak,
winning over the negative ones from the cross contributions. It is
important to note that in the alkyl tail carbon atom contributions,
those carbons close to the hydroxyl head contribute to a positive
pre-peak (since they are strongly attached to the head group), while
the cross correlation of these atoms with the trailing carbons contribute
to a negative anti-peak. This explains why there are negative contributions
in the gray curves as one goes to longer alcohols. It is also noteworthy
that structure factors between trailing carbon atoms do not have any
pre-peak, and look just standard Lennard-Jones structure factors.
In other words, the tail atoms by themselves do not contribute to
the pre-peak. This remark shows the importance of interpreting $I(k)$
as resulting from statistical correlations between different parts
of the meta-objects, and not only to some of them, such as the hydroxyl
group chain. 

Looking closer at the lower plots, we notice that the structure factors
of ethanol have almost no anti-peaks. The reason for this is simply
because there is only one neutral atom: the tailing methyl group.
This is not the case any more for 1-pentanol and 1-octanol, and we
clearly see the importance of the anti-peaks. The positive pre-peak
of the hydroxyl correlations also grows in intensity when going from
ethanol to 1-octanol, reflecting the tightness of the hydrogen bonding
for larger mono-ols. Another feature is the apparent inversion of
peak heights between the structure factors and $I(k)$. Indeed, the
structure factors show a large pre-peak and a smaller main peak, while
this is inverted in $I(k)$. The reason is that it is the number of
contributions in Eq.(\ref{Ik}) which matters, and the smaller main
peaks of the structure factor end up overwhelming that of the pre-peak,
resulting in the actual shape of $I(k)$. Details of these correlations,
intentionally undifferentiated here, will be discussed in a subsequent
work.

\subsubsection{The important role of the intra-molecular correlations}

In the previous part we have seen that the cross domain correlations
involve large negative contributions in $I(k)$. Such contributions
could potentially cancel the pre-peak contributions in some cases,
and they indeed do, as we have demonstrated for the case of aqueous
1-propanol mixture \cite{AupDomainOrdering}. This observation poses
the problem of how to interpret the appearance of a pre-peak in $I(k)$
as residual part of canceling effects, and more importantly how to
relate this to an underlying microscopic physics such as self-assembly.
For example, in another work \cite{AupPropylamine2}, we interpreted
the experimental evidence for the appearance of a scattering pre-peak
in aqueous 1-propylamine mixtures, as opposed to its absence in aqueous
1-propanol mixtures, as the signature of water hydrogen bonding to
the amine and hydroxyl groups is different for each type of mixtures.
In the present case, we observe that, if instead of the correct $w_{ij}(k)$
involving the molecular flexibility, we use the approximation of the
rigid molecule (see Section \ref{subsec:The-intra-molecular-correlations}),
then the pre-peak of longer alcohols vanishes, or even becomes a negative
anti-peak, which is an unphysical feature since $I(k)$ should be
positive for all $k$. This is illustrated in Fig.12, where we show,
for the case of the 3 alcohols selected for Fig.10, the like $I_{\mbox{like}}(k)$
and cross contributions $I_{\mbox{cross}}(k)$ together with the total
$I(k)$, for the cases of the rigid and flexible $w_{ij}(k)$, respectively
in dashed and full lines.

\begin{figure}[ht]
\centering \includegraphics[height=4.5cm]{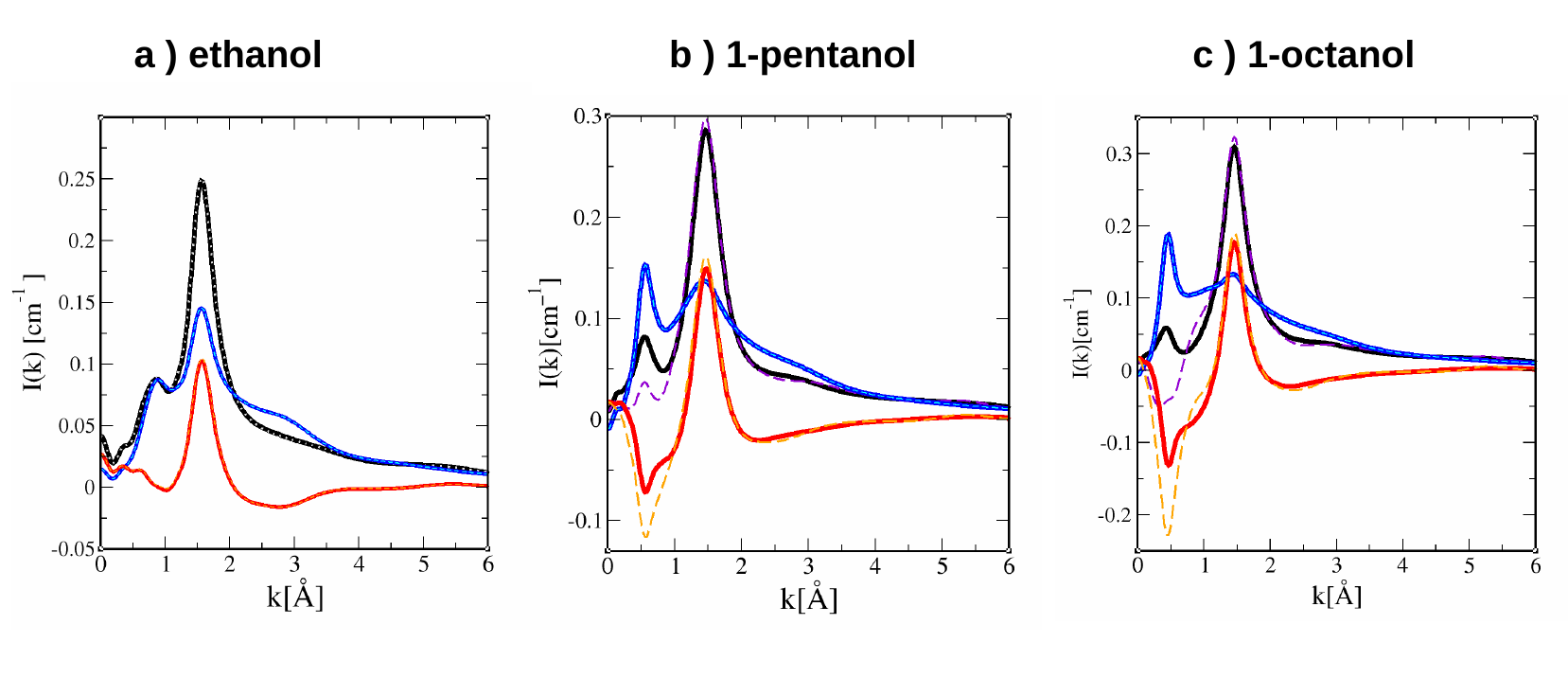} \label{fig13-IkPart}
\caption{Illustration of the crucial influence of flexibility in the pre-peak
description, for the OPLS model of 3 alcohols (ethanol, 1-pentanol
and 1-octanol). The full curves are for when flexibility is properly
included in the $w_{ij}(k)$ functions (see Fig.1), and thin dashed
curves when $w_{ij}(k)$ for rigid molecules is used. Black curves
are for total $I(k)$, blue curve for like atom-contributions of $I(k)$
and red curves for cross atom contributions.}
\end{figure}

It is quite clear that, in the case of the rigid intra-molecular approximation
with dashed lines, the pre-peak feature is considerably modified and
does not match the respective experimental data. Also, this effect
is more pronounced for longer alkanols. Indeed, for 1-octanol, $I(k)$
is seen to develop an unphysical negative anti-peak (black dashed
curve in the left panel) when flexibility is not properly described,
and the dashed red curve shows that the negative contribution comes
from the overwhelming cross atoms structure factor, involving in particular
correlations between the hydroxyl h

\section{Discussion\label{sec:Discussion}}

Radiation scattering in liquids is a statistical probe of the microscopic
features in the disorder, which is why it is related to purely statistical
observable such as correlation functions and structure factors, as
in the Debye formula. It is the analysis of the structure factors
which would uncover which microscopic features contribute to specific
features of the scattered intensity, such as pre-peaks. Many types
of investigations of alcohols point towards the existence of hydroxyl
group aggregates. But the present work shows clearly that it is not
only these aggregates which contribute to the features of $I(k$),
but the entire meta-object formed by the surrounding alkyl tails as
well, which is seen by the scattered radiation. The alkyl tails are
shown to play an increasingly important role with $n$, particularly
through their flexibility. More importantly, it appears necessary
to include the anti-correlations between the hydroxyl head group and
the alkyl tail, in order to interpret the decrease of the pre-peak
when these alkyl tails grow in size beyond $n=4$. The study clearly
suggests that the pre-peak cannot be fully understood with a model
free investigation, hence necessarily involving the penalties associated
with modeling shortcomings.

The comparison of the experimental and simulated $I(k)$ shows the
strong model dependence, and in particular the erratic behaviour of
the CHARMM model deserves some investigation. This model differs from
the OPLS and TraPPE models by the all-atom representation of the methylene
and methyl groups, but more importantly by the fact that the corresponding
hydrogen and central carbon bear relatively large partial charges.
In addition, unlike the two other models which assign the same partial
charges to the same atoms throughout the various alcohols, the CHARMM
model has different adjustments depending on alcohols (see Tables).
In view of the argumentation in the previous section, where we have
shown the importance of charge ordering in the microscopic structure,
we believe that this model blurs the charge ordering and diminish
strongly the hydroxyl group association.

It is an experimental observation that the pre-peak in $I(k)$ is
always positive. In view of Eq.(\ref{Ik}) and the existence of pre-peak
and anti-peaks in the $S_{ij}(k)$, one may ask why their summed contribution
in Eq.(\ref{Ik}) should always lead to a positive contribution in
the experiments. The example of the CHARMM model shows that this is
not always the case, as in the case of 1-pentanol, hence indicating
that this model has something unphysical about it. The CHARMM model,
in a sense, demonstrates that it is possible to obtain a micro-structure
obeying charge ordering and domain ordering criteria, and yet give
unphysical quantities, such as no-prepeak or negative anti-peaks.

However, one should not conclude that all all-atom models are bound
to such fate. Indeed, we have tested the OPLS-AA for methanol and
ethanol, and it gives a pre-peak for $I(k)$, although somewhat lower
than expected, when compared to the OPLS-UA model. Looking at the
partial charges of this model in the Tables, we notice that those
on the methyl/methylene groups are somewhat smaller than those of
the CHARMM model. In a way, these groups will have less tendency to
blur the charge ordering.

In order to support these arguments, we show in Fig.13 a comparison
of the OPLS-AA model for methanol and ethanol with OPLS-UA and CHARMM.

\begin{figure}[ht]
\centering \includegraphics[height=6cm]{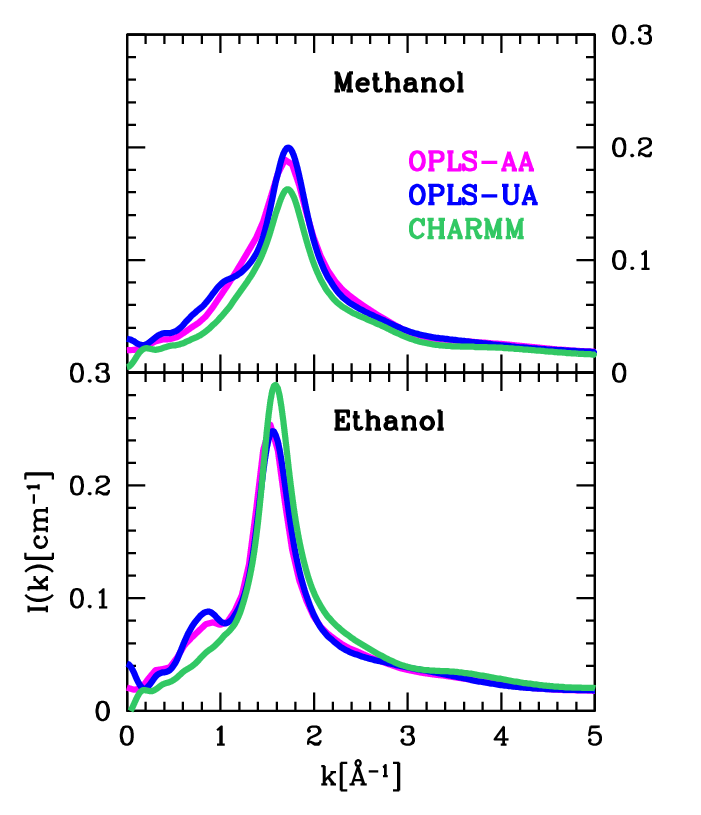} \label{fig14-Ikmodel}
\caption{Calculated $I(k)$ for methanol(top panel) and ethanol(bottom panel)
illustrating the ability of all-atom and united-atom models to reproduce
the pre-peak. CHARMM-AA in green, OPLS-AA in magenta and OPLS-UA in
green.}
\end{figure}

It is seen that, for methanol, the OPLS-AA model misses the pre-peak
shoulder feature, just like the CHARMM model. For ethanol however,
OPLS-AA shows a pre-peak, although weaker than the OPLS-UA model.
When looking at the charges of the terminal C group in Table 5, we
find that OPLS-AA has 0.06 for the hydrogen, while CHARM has 0.09,
a higher value. In other words, the Coulomb interaction of the carbon
group hydrogen would be more strongly felt for CHARMM than for OPLS-AA,
leading to a stronger charge order homogeneity breaking for this latter
model. This explains why OPLS-AA has stronger pre-peak than CHARMM.

Interestingly, OPLS-AA has a cluster distribution similar to that of
CHARMM in Fig.7, suggesting that all AA models tend to favor higher
monomer probability than their UA counterpart. We note that this is
not in contradiction with the similarity of the correlation functions
and structure factors in what concerns the clustering features, since
AA models have also a large number of clusters, as indicated by the
large cluster distribution tail in Fig.6 for CHARMM.

\section{Conclusion}

In this work, we have compared the calculated X-ray scattering intensities
$I(k)$ for a variety of mono-ols, ranging from methanol to 1-nonanol,
to the corresponding experimental data (methanol to 1-undecanol),
and for several of known force field models, such as OPLS, TraPPE,
CHARMM and GROMOS. The principal focus of this calculation was twofold:
on one hand to test the ability of various types of force field models
to reproduce the overall shape of the experimental $I(k)$, and particularly 
the well known scattering pre-peak feature of these alcohols, and
on the other hand to provide an explanation for the generic features
of $I(k)$ across different alcohols. Two notable such features are the
decrease of $k_{P}$ with increasing values of $n$, and the amplitude
$A_{P}$ turnover around 1-pentanol. 

As far as the first point is concerned, we find in general that the
OPLS-UA model provides better results. In particular it 
reproduces well the pre-peak's position and relative intensity
variations throughout the whole series of mono-ols studied. The TraPPE
model is less performant, particularly in what concerns the main-peak.
But this model predicts pre-peak for all alcohols. The CHARMM model
is the most problematic, particularly for small alcohols, since it
fails to predict the pre-peak for methanol and ethanol, and predicts
an unphysical negative anti peak for 1-pentanol. However, it is good
for longer alcohols. The problems found for TraPPE and CHARMM around
1-pentanol could well be related to the physical fact that there is
a crossover of behaviour of the aggregated structures around that
particular alcohol.
\begin{figure}[ht]
\centering \includegraphics[height=6cm]{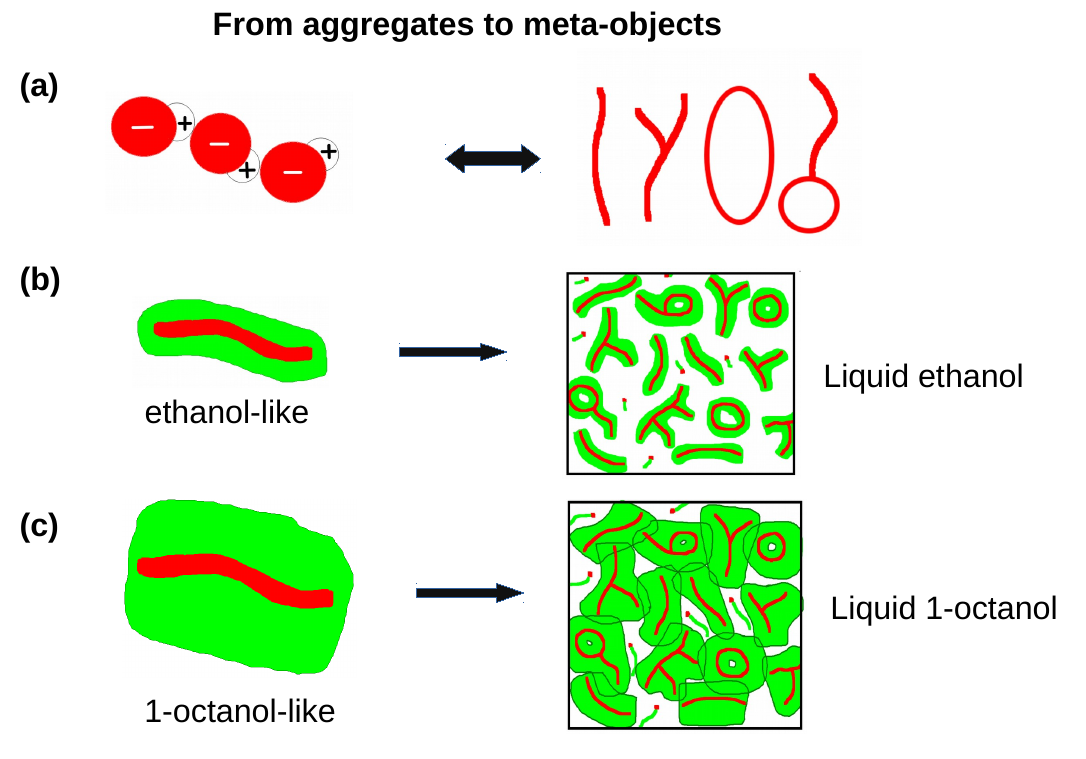} \label{fig15-GCO}
\caption{Illustration of the concept of a liquid of meta-objects, starting from
 the initial charge ordered chain (a) with alternating O(-) and H(+) hydroxyl
 heads, generating various shapes of clusters.
  Meta-object representation of ethanol (b) and 1-octanol (c) with the central chain (red) surrounded by alkyl tail cloud (in green) with few monomers. }
\end{figure}
As far as the second point is concerned, we have shown that the origin
of the pre-peak in $I(k)$ is not only related to the hydroxyl head
groups aggregates, but to the entire meta-object formed by this central
chain and the surrounding alkyl tails, promoting the picture of such
alcohols made not only of monomer alcohol molecules, but also meta-objects.
This is illustrated in Fig.15.
$I(k)$ would then detect both the atomic constituents through the
main peak and the meta-object through the pre-peak. More specifically,
the pre-peak is the result of correlations between the charged head
groups part and cross-correlations between the head and tail parts.
The duality of the meta-object and the correlations between its part
appears as a whole. Earliser studies of methanol reported the large
variety of cluster shapes, such as chains, simple and branched,
loops, lassos, etc...
\cite{ExpScattMaginiMethanol,ExpScattNartenEthMeth,EXP_Sarkar_Joarder_Methanol,
EXP_Sarkar_Joarder_ethanol,EXP_SANS_Soper_pure_methanol,ExpScattJoarderAlcohols,
EXP_Joarder_Methanol_temp,ExpScattBenmoreEthanol,EXP_Pustai_MethEthProp,
AUP_Neat_Alcohols_JCP,AUP_Neat_alcohols_PRE}.
The present work indicates that this is also the case for all mono-ols.

Pre-peaks are found in many soft-matter systems \cite{EXP_Teubner_Strey,
EXP_Microemulsions1,EXP_Microemulsions2,Buening2017,Bierwirth2018},
particularly in micro-emulsion systems, and geometric shapes of the
aggregates are often deduced from the shape of $I(k)$. In view of
Eq.(\ref{Iraw}) and the fact that aggregates are better defined from
the atomic structure factors themselves, it is quite possible that
a more detailed analysis and interpretation in terms of charge and
domain order via atom-atom correlations may provide deeper insight
into self-assembly. In any case, the need for computer simulations
appears as necessary, in order to calculate the atom-atom structure
factors. This could pose quite formidable modeling problems, specially
when considering typical soft-matter systems. We hope that the present
investigation has provided elements for reconsidering usual interpretations
of $I(k)$ in terms of microscopic properties, by establishing the
need for the dual aggregate/correlation picture.

\section*{Acknowledgments}

This work has been supported in part by the Croatian Science Foundation
under the project UIP-2017-05-1863 ``Dynamics in micro-segregated
systems''. We thank the DELTA machine group for providing synchrotron
radiation and technical support, Michael Paulus for help with the
X-ray diffraction experiments at beamlines BL2 and BL9, and Metin
Tolan for discussion and supporting these activities. We would also
like to thank Lena Friedrich and Pia-Victoria Pottkämper for their
help with the measurements.


\begin{thebibliography}{10}

\bibitem{ExpScattOldAlc2}
Warren, B.
\newblock X-ray diffraction in long chain liquids.
\newblock \emph{Phys. Rev.}, \textbf{1933}.
\newblock \emph{44}, 969.

\bibitem{ExpScattOldAlc}
Pierce, W. and MacMillan, D.
\newblock X-ray studies on liquids: the inner peak for alcohols and acids.
\newblock \emph{Journal of the American Chemical Society}, \textbf{1938}.
\newblock \emph{60}, 779.

\bibitem{EXP_Kruh_diffraction}
Kruh, R.F.
\newblock Diffraction studies of the structure of liquids.
\newblock \emph{Chemical Reviews}, \textbf{1962}.
\newblock \emph{62}(4), 319.

\bibitem{EXP_Sarkar_Joarder_Methanol}
Sarkar, S. and Joarder, R.N.
\newblock Molecular clusters and correlations in liquid methanol at room
  temperature.
\newblock \emph{The Journal of Chemical Physics}, \textbf{1993}.
\newblock \emph{99}, 2032.

\bibitem{EXP_Sarkar_Joarder_ethanol}
Sarkar, S. and Joarder, R.N.
\newblock Molecular clusters in liquid ethanol at room temperature.
\newblock \emph{The Journal of Chemical Physics}, \textbf{1994}.
\newblock \emph{100}, 5118.

\bibitem{Marcus_Water_Octanol}
Marcus, Y.
\newblock Structural aspects of water in 1-octanol.
\newblock \emph{Journal of Solution Chemistry}, \textbf{1990}.
\newblock \emph{19}, 507 .

\bibitem{Frank_Abraham_Lieb_Octanol_Hard_sphere}
Franks, N.P.; Abraham, M.H. and Lieb, W.R.
\newblock Molecular organization of liquid n-octanol: An x-ray diffraction
  analysis.
\newblock \emph{Journal of Pharmaceutical Sciences}, \textbf{1993}.
\newblock \emph{82}(5), 466.

\bibitem{SimulationLudwigMethanol}
Ludwig, R.
\newblock The structure of liquid methanol.
\newblock \emph{ChemPhysChem}, \textbf{2005}.
\newblock \emph{6}, 1369.

\bibitem{AUP_Neat_Alcohols_JCP}
Zorani\'{c}, L.; Sokoli\'{c}, F. and Perera, A.
\newblock Microstructure of neat alcohols: A molecular dynamics study.
\newblock \emph{The Journal of Chemical Physics}, \textbf{2007}.
\newblock \emph{127}, 024502.

\bibitem{AUP_Neat_alcohols_PRE}
Perera, A.; Sokoli\'{c}, F. and Zorani\'{c}, L.
\newblock Microstructure of neat alcohols.
\newblock \emph{Physical Review E}, \textbf{2007}.
\newblock \emph{75}, 060502(R).

\bibitem{SIM_Finci2_linear_alcohols}
Lehtola, J.; Hakala, M. and H\"{a}m\"{a}l\"{a}inen, K.
\newblock Structure of liquid linear alcohols.
\newblock \emph{The Journal of Physical Chemistry B}, \textbf{2010}.
\newblock \emph{114}, 6426.

\bibitem{EXP_Pustai_MethEthProp}
Vrhov\v{s}ek, A.; Gereben, O.; Jamnik, A. and Pusztai, L.
\newblock Hydrogen bonding and molecular aggregates in liquid methanol,
  ethanol, and 1-propanol.
\newblock \emph{The Journal of Physical Chemistry B}, \textbf{2011}.
\newblock \emph{115}, 13473.

\bibitem{SIM_MacCallum_Octanol}
MacCallum, J.L. and Tieleman, D.P.
\newblock Structures of neat and hydrated 1-octanol from computer simulations.
\newblock \emph{Journal of the American Chemical Society}, \textbf{2002}.
\newblock \emph{124}(50), 15085.
\newblock PMID: 12475354.

\bibitem{SIM_Siepmann_Octanol}
Chen, B. and Siepmann, J.I.
\newblock Microscopic structure and solvation in dry and wet octanol.
\newblock \emph{The Journal of Physical Chemistry B}, \textbf{2006}.
\newblock \emph{110}(8), 3555.
\newblock PMID: 16494411.

\bibitem{SIM_Mariani_alcohols_pressure}
Mariani, A.; Ballirano, P.; Angiolari, F.; Caminiti, R. and Gontrani, L.
\newblock Does high pressure induce structural reorganization in linear
  alcohols? a computational answer.
\newblock \emph{ChemPhysChem}, \textbf{2016}.
\newblock \emph{17}(19), 3023.

\bibitem{Tomsic_butanol}
Cerar, J.; Lajovic, A.; Jamnik, A. and Tom\v{s}i\v{c}, M.
\newblock Performance of various models in structural characterization of
  n-butanol: Molecular dynamics and x-ray scattering studies.
\newblock \emph{Journal of Molecular Liquids}, \textbf{2017}.
\newblock \emph{229}, 346 .

\bibitem{Nano_reviews_Boldon}
Boldon, L.; Laliberte, F. and Liu, L.
\newblock Review of the fundamental theories behind small angle x-ray
  scattering, molecular dynamics simulations, and relevant integrated
  application.
\newblock \emph{Nano Reviews}, \textbf{2015}.
\newblock \emph{6}(1), 25661.
\newblock PMID: 25721341.

\bibitem{Debye1}
Debye, P.
\newblock Zerstreuung von röntgenstrahlen.
\newblock \emph{Annalen der Physik}, \textbf{1915}.
\newblock \emph{351}, 809.

\bibitem{Debye2}
Debye, P.
\newblock Scattering of x-rays.
\newblock In \emph{The collected papers of Peter J.W. Debye}. Interscience
  Publishers, \textbf{1954}.

\bibitem{SimulationBakoNeatMethClusters}
Bak\'{o}, I.; Jedlovszky, P. and P\'{a}link\'{a}s, G.
\newblock Molecular clusters in liquid methanol: a reverse monte carlo study.
\newblock \emph{Journal of Molecular Liquids}, \textbf{2000}.
\newblock \emph{87}, 243.

\bibitem{ExpScattBenmoreEthanol}
Benmore, C. and Loh, Y.
\newblock The structure of liquid ethanol: A neutron diffraction and molecular
  dynamics study.
\newblock \emph{The Journal of Chemical Physics}, \textbf{2000}.
\newblock \emph{112}, 5877.

\bibitem{SIM_Bako_methanol_EXP}
Kosztol\'{a}nyi, T.; Bak\'{o}, I. and P\'{a}link\'{a}s, G.
\newblock Hydrogen bonding in liquid methanol, methylamine, and methanethiol
  studied by molecular-dynamics simulations.
\newblock \emph{The Journal of Chemical Physics}, \textbf{2003}.
\newblock \emph{118}, 4546.

\bibitem{ExpScattMatijaMonools}
Tom\v{s}i\v{c}, M.; Jamnik, A.; Fritz-Popovski, G.; Glatter, O. and Vl\v{c}ek,
  L.
\newblock Structural properties of pure simple alcohols from ethanol, propanol,
  butanol, pentanol, to hexanol: Comparing monte carlo simulations with
  experimental saxs data.
\newblock \emph{The Journal of Physical Chemistry B}, \textbf{2007}.
\newblock \emph{111}, 1738.

\bibitem{Ionic_1}
Santos, C.; Annapureddy, H.R.; Murthy, N.; Kashyap, H.; Castner, E. and
  Margulis, C.
\newblock Temperature-dependent structure of methyltributylammonium
  bis(trifluoromethylsulfonyl)amide: X ray scattering and simulations.
\newblock \emph{The Journal of Chemical Physics}, \textbf{2011}.
\newblock \emph{134}(6), 064501.

\bibitem{Ionic_2}
Siqueira, L. and Ribeiro, M.
\newblock Charge ordering and intermediate range order in ammonium ionic
  liquids.
\newblock \emph{The Journal of Chemical Physics}, \textbf{2011}.
\newblock \emph{135}(20), 204506.

\bibitem{Ionic_3}
Annapureddy, H.; Kashyap, H.; De~Biase, P. and Margulis, C.
\newblock What is the origin of the prepeak in the x-ray scattering of
  imidazolium-based room-temperature ionic liquids?
\newblock \emph{The Journal of Physical Chemistry B}, \textbf{2010}.
\newblock \emph{114}(50), 16838.
\newblock PMID: 21077649.

\bibitem{Triolo_Amphiphile}
Russina, O.; Sferrazza, A.; Caminiti, R. and Triolo, A.
\newblock Amphiphile meets amphiphile: Beyond the polar-apolar 
dualism in ionic liquid alcohol mixtures.
\newblock \emph{The Journal of Physical Chemistry Letters}, \textbf{2014}.
\newblock \emph{5}(10), 1738.
\newblock PMID: 26270376.

\bibitem{krywka2007small}
Krywka, C.; Sternemann, C.; Paulus, M.; Javid, N.; Winter, R.; Al-Sawalmih, A.;
  Yi, S.; Raabe, D. and Tolan, M.
\newblock The small-angle and wide-angle x-ray scattering set-up at beamline
  bl9 of delta.
\newblock \emph{Journal of synchrotron radiation}, \textbf{2007}.
\newblock \emph{14}(3), 244.

\bibitem{hammersley1996two}
Hammersley, A.; Svensson, S.; Hanfland, M.; Fitch, A. and Hausermann, D.
\newblock Two-dimensional detector software: from real detector to idealised
  image or two-theta scan.
\newblock \emph{International Journal of High Pressure Research},
  \textbf{1996}.
\newblock \emph{14}(4-6), 235.

\bibitem{FF_OPLS_alcohols_1}
Jorgensen, W.
\newblock Optimized intermolecular potential functions for liquid alcohols.
\newblock \emph{The Journal of Physical Chemistry}, \textbf{1986}.
\newblock \emph{90}, 1276.

\bibitem{FF_OPLS_AA_organic_liquids_amines}
Jorgensen, W.; Maxwell, D. and Tirado-Rives, J.
\newblock Development and testing of the opls all-atom force field on
  conformational energetics and properties of organic liquids.
\newblock \emph{Journal of the American Chemical Society}, \textbf{1996}.
\newblock \emph{118}, 11225.

\bibitem{FF_Trappe_Alcohols}
Chen, B.; Potoff, J. and Siepmann, J.
\newblock Monte carlo calculations for alcohols and their mixtures with
  alkanes. transferable potentials for phase equilibria. 5. united-atom
  description of primary, secondary, and tertiary alcohols.
\newblock \emph{The Journal of Physical Chemistry B}, \textbf{2001}.
\newblock \emph{105}, 3093.

\bibitem{FF_Charmm1}
Vanommeslaeghe, K.; Hatcher, E.; Acharya, C.; Kundu, S.; Zhong, S.; Shim, J.;
  Darian, E.; Guvench, O.; Lopes, P.; Vorobyov, I. and Mackerell~Jr., A.D.
\newblock Charmm general force field: A force field for drug-like molecules
  compatible with the charmm all-atom additive biological force fields.
\newblock \emph{Journal of Computational Chemistry}, \textbf{2010}.
\newblock \emph{31}, 671.

\bibitem{FF_Charmm2}
Vanommeslaeghe, K. and MacKerell, A.D.
\newblock Automation of the charmm general force field (cgenff) i: Bond
  perception and atom typing.
\newblock \emph{Journal of Chemical Information and Modeling}, \textbf{2012}.
\newblock \emph{52}, 3144.

\bibitem{FF_Charmm3}
Vanommeslaeghe, K.; Raman, E.P. and MacKerell, A.D.
\newblock Automation of the charmm general force field (cgenff) ii: Assignment
  of bonded parameters and partial atomic charges.
\newblock \emph{Journal of Chemical Information and Modeling}, \textbf{2012}.
\newblock \emph{52}, 3155.

\bibitem{FF_Gromos_54a7}
Schmid, N.; Eichenberger, A.; Choutko, A.; Riniker, S.; Winger, M.; Mark, A.
  and van Gunsteren, W.
\newblock Definition and testing of the gromos force-field versions 54a7 and
  54b7.
\newblock \emph{European Biophysics Journal}, \textbf{2011}.
\newblock \emph{40}, 843.

\bibitem{FF_Gromos_ATB}
Malde, A.K.; Zuo, L.; Breeze, M.; Stroet, M.; Poger, D.; Nair, P.C.;
  Oostenbrink, C. and Mark, A.E.
\newblock An automated force field topology builder (atb) and repository:
  Version 1.0.
\newblock \emph{Journal of Chemical Theory and Computation}, \textbf{2011}.
\newblock \emph{7}, 4026.

\bibitem{SimulationGromacs}
Pronk, S.; P\'{a}ll, S.; Schulz, R.; Larsson, P.; Bjelkmar, P.; Apostolov, R.;
  Shirts, M.; Smith, J.; Kasson, P.; van der Spoel, D.; Hess, B. and Lindahl,
  E.
\newblock Gromacs 4.5: a high-throughput and highly parallel open source
  molecular simulation toolkit.
\newblock \emph{Bioinformatics}, \textbf{2013}.
\newblock \emph{29}, 845.

\bibitem{MD_thermo_Nose}
Nose, S.
\newblock A molecular dynamics method for simulations in the canonical
  ensemble.
\newblock \emph{Molecular Physics}, \textbf{1984}.
\newblock \emph{52}, 255.

\bibitem{MD_thermo_Hoover}
Hoover, W.
\newblock Canonical dynamics: Equilibrium phase-space distributions.
\newblock \emph{Physical Review A}, \textbf{1985}.
\newblock \emph{31}, 1695.

\bibitem{MD_barostat_Parrinello_Rahman_1}
Parrinello, M. and Rahman, A.
\newblock Crystal structure and pair potentials: A molecular-dynamics study.
\newblock \emph{Physical Review Letters}, \textbf{1980}.
\newblock \emph{45}, 1196.

\bibitem{MD_barostat_Parrinello_Rahman_2}
Parrinello, M. and Rahman, A.
\newblock Polymorphic transitions in single crystals: A new molecular dynamics
  method.
\newblock \emph{Journal of Applied Physics}, \textbf{1981}.
\newblock \emph{52}, 7182.

\bibitem{MD_Packmol}
Mart\'{i}nez, J. and Mart\'{i}nez, L.
\newblock Packing optimization for automated generation of complex system's
  initial configurations for molecular dynamics and docking.
\newblock \emph{Journal of Computational Chemistry}, \textbf{2003}.
\newblock \emph{24}, 819.

\bibitem{Textbook_Hansen_McDonald}
Hansen, J.P. and McDonald, I.
\newblock \emph{Theory of Simple Liquids}.
\newblock Academic Press, Elsevier, Amsterdam, 3rd edition, \textbf{2006}.

\bibitem{2015_PCCP_benzMH}
Po\v{z}ar, M.; Seguier, J.B.; Guerche, J.; Mazighi, R.; Zorani\'{c}, L.;
  Mijakovi\'{c}, M.; Ke\v{z}i\'{c}-Lovrin\v{c}evi\'{c}, B.; Sokoli\'{c}, F. and
  Perera, A.
\newblock Simple and complex disorder in binary mixtures with benzene as a
  common solvent.
\newblock \emph{Physical Chemistry Chemical Physics}, \textbf{2015}.
\newblock \emph{17}, 9885.

\bibitem{2016_PCCP_MH_Versus_Clust}
Po\v{z}ar, M.; Lovrin\v{c}evi\'{c}, B.; Zorani\'{c}, L.; Primorac, T.;
  Sokoli\'{c}, F. and Perera, A.
\newblock Micro-heterogeneity versus clustering in binary mixtures of ethanol
  with water or alkanes.
\newblock \emph{Physical Chemistry Chemical Physics}, \textbf{2016}.
\newblock \emph{18}, 23971.

\bibitem{AUP_Charge_ordering_prepeak_neat_alc}
Perera, A.
\newblock Charge ordering and scattering pre-peaks in ionic liquids and
  alcohols.
\newblock \emph{Physical Chemistry Chemical Physics}, \textbf{2017}.
\newblock \emph{19}, 1062.

\bibitem{AupPropylamine2}
Alm\'{a}sy, L.; Kuklin, A.; Po\v{z}ar, M.; Baptista, A. and Perera, A.
\newblock Microscopic origin of the scattering pre-peak in aqueous propylamine
  mixtures: X-ray and neutron experiments versus simulations.
\newblock \emph{Phys. Chem. Chem. Phys.}, \textbf{2019}.
\newblock \emph{21}, 9317.

\bibitem{ExpScattFinnsMonools}
Vahvaselk\"{a}, K.S.; Serimaa, R. and Torkkeli, M.
\newblock Determination of liquid structures of the primary alcohols methanol,
  ethanol, 1-propanol, 1-butanol and 1-octanol by x-ray scattering.
\newblock \emph{Journal of Applied Crystallography}, \textbf{1995}.
\newblock \emph{28}, 189.

\bibitem{ExpScattMaginiMethanol}
Magini, M.; Paschina, G. and Piccaluga, G.
\newblock On the structure of methyl alcohol at room temperature.
\newblock \emph{The Journal of Chemical Physics}, \textbf{1982}.
\newblock \emph{77}, 2051.

\bibitem{ExpScattNartenEthMeth}
Narten, A. and Habenschuss, A.
\newblock Hydrogen bonding in liquid methanol and ethanol determined by x-ray
  diffraction.
\newblock \emph{The Journal of Chemical Physics}, \textbf{1984}.
\newblock \emph{80}, 3387.

\bibitem{ExpScattJoarderAlcohols}
Karmakar, A.; Krishna, P. and Joarder, R.
\newblock On the structure function of liquid alcohols at small wave numbers
  and signature of hydrogen-bonded clusters in the liquid state.
\newblock \emph{Physics Letters A}, \textbf{1999}.
\newblock \emph{253}, 207.

\bibitem{EXP_Sarkar_1propanol}
Sahoo, A.; Sarkar, S.; Bhagat, V. and Joarder, R.N.
\newblock The probable molecular association in liquid d-1-propanol through
  neutron diffraction.
\newblock \emph{The Journal of Physical Chemistry A}, \textbf{2009}.
\newblock \emph{113}, 5160.

\bibitem{EXP_Joarder_Methanol_temp}
Sahoo, A.; Nath, P.; Bhagat, V.; Krishna, P. and Joarder, R.
\newblock Effect of temperature on the molecular association in liquid
  d-methanol using neutron diffraction data.
\newblock \emph{Physics and Chemistry of Liquids}, \textbf{2010}.
\newblock \emph{48}, 546.

\bibitem{EXP_Silleren_Propanol}
Sillr\'{e}n, P.; Swenson, J.; Mattsson, J.; Bowron, D. and Matic, A.
\newblock The temperature dependent structure of liquid 1-propanol as studied
  by neutron diffraction and epsr simulations.
\newblock \emph{The Journal of Chemical Physics}, \textbf{2013}.
\newblock \emph{138}, 214501.

\bibitem{Tomsic_Brij35_Water_Alcohols}
Tom\v{s}i\v{c}, M.; Be\v{s}ter-Roga\v{c}, M.; Jamnik, A.; Kunz, W.; Touraud,
  D.; Bergmann, A. and Glatter, O.
\newblock Nonionic surfactant brij 35 in water and in various simple alcohols: 
structural investigations by small angle x-ray scattering and dynamic light scattering.
\newblock \emph{The Journal of Physical Chemistry B}, \textbf{2004}.
\newblock \emph{108}(22), 7021.

\bibitem{Lara_Ethanol_MolSim}
Mijakovi\'{c}, M.; Polok, K.; Ke\v{z}i\'{c}, B.; Sokoli\'{c}, F.; Perera, A.
  and Zorani\'{c}, L.
\newblock A comparison of force fields for ethanol-water mixtures.
\newblock \emph{Molecular Simulation}, \textbf{2015}.
\newblock \emph{41}(9), 699.

\bibitem{2016_JCP_ethMeth}
Po\v{z}ar, M.; Lovrin\v{c}evi\'{c}, B.; Zorani\'{c}, L.; Mijakovi\'{c}, M.;
  Sokoli\'{c}, F. and Perera, A.
\newblock A re-appraisal of the concept of ideal mixtures through a computer
  simulation study of the methanol-ethanol mixtures.
\newblock \emph{The Journal of Chemical Physics}, \textbf{2016}.
\newblock \emph{145}, 064509.

\bibitem{AupPAC}
Perera, A.
\newblock From solutions to molecular emulsions.
\newblock \emph{Pure and Applied Chemistry}, \textbf{2016}.
\newblock \emph{88}, 189.

\bibitem{AupDomainOrdering}
Perera, A.
\newblock Molecular emulsions: from charge order to domain order.
\newblock \emph{Phys. Chem. Chem. Phys.}, \textbf{2017}.
\newblock \emph{19}, 28275.

\bibitem{EXP_SANS_Soper_pure_methanol}
Yamaguchi, T.; Hidaka, K. and Soper, A.
\newblock The structure of liquid methanol revisited: a neutron diffraction
  experiment at -80c and 25c.
\newblock \emph{Molecular Physics}, \textbf{1999}.
\newblock \emph{96}, 1159.

\bibitem{EXP_Teubner_Strey}
Teubner, M. and Strey, R.
\newblock Origin of the scattering peak in microemulsions.
\newblock \emph{The Journal of Chemical Physics}, \textbf{1987}.
\newblock \emph{87}(5), 3195.

\bibitem{EXP_Microemulsions1}
Goddeeris, C.; Cuppo, F.; Reynaers, H.; Bouwman, W. and den Mooter, G.V.
\newblock Light scattering measurements on microemulsions: Estimation of
  droplet sizes.
\newblock \emph{International Journal of Pharmaceutics}, \textbf{2006}.
\newblock \emph{312}, 187 .

\bibitem{EXP_Microemulsions2}
Pr\'{e}vost, S.; Gradzielski, M. and Zemb, T.
\newblock Self-assembly, phase behaviour and structural behaviour as observed
  by scattering for classical and non-classical microemulsions.
\newblock \emph{Advances in Colloid and Interface Science}, \textbf{2017}.
\newblock \emph{247}, 374 .

\bibitem{Buening2017}
B\"{u}ning, T.; Lueg, J.; Bolle, J.; Sternemann, C.; Gainaru, C.; Tolan, M. and
  B\"{o}hmer, R.
\newblock Connecting structurally and dynamically detected signatures of
  supramolecular debye liquids.
\newblock \emph{The Journal of Chemical Physics}, \textbf{2017}.
\newblock \emph{147}(23), 234501.

\bibitem{Bierwirth2018}
Bierwirth, S.P.; Bolle, J.; Bauer, S.; Sternemann, C.; Gainaru, C.; Tolan, M.
  and B{\"o}hmer, R.
\newblock \emph{Scaling of Suprastructure and Dynamics in Pure and Mixed Debye
  Liquids}, pages 121--171.
\newblock Springer International Publishing, Cham, \textbf{2018}.

\end{thebibliography}

\end{document}